\documentclass[12pt]{article}
\usepackage{a4}\usepackage{amsfonts}\textwidth15cm
\usepackage{graphicx}
\newcommand{\Ref}[1]{(\ref{#1})}
\newcommand{\be}{\begin{equation}}\newcommand{\ee}{\end{equation}}
\newcommand{\bea}{\begin{eqnarray}}\newcommand{\eea}{\end{eqnarray}}
\newcommand{\beq}{\begin{eqnarray}}\newcommand{\eeq}{\end{eqnarray}}
\newcommand{\beao}{\begin{eqnarray*}}\newcommand{\eeao}{\end{eqnarray*}}
\renewcommand{\S}{{\cal S}}
  \newcommand{\A}{{\cal K}}
\newcommand{\M}{{\cal M}}
\newcommand{\nn}{\nonumber}\newcommand{\pa}{\partial}
\newcommand{\al}{\alpha}\newcommand{\ep}{\epsilon}
\newcommand{\la}{\lambda}\newcommand{\ga}{\gamma}
\newcommand{\om}{\omega}\newcommand{\Om}{\Omega}
\newcommand{\Li}{{\rm Li}}
\newcommand{ \text}[1]{#1 }
\newcommand{\meins}{\cite{Bordag:2006vc}}
\begin{document}
\title{Generalized Lifshitz formula for a cylindrical plasma sheet in front of a plane beyond proximity force approximation}
\author{
{\sc M. Bordag}\thanks{e-mail: Michael.Bordag@itp.uni-leipzig.de} \\
\small  University of Leipzig, Institute for Theoretical Physics\\
\small  Vor dem Hospitaltore 1, 04103 Leipzig, Germany}
\maketitle
\begin{abstract}
We calculate  the first correction beyond proximity force approximation for a cylindrical graphene sheet in interaction with a flat graphene sheet or a  dielectric half space.
\end{abstract} 
\section{Introduction}
The interaction of material bodies at distances in the micrometer and nanometer  scale is of significant actual interest in view of applications in nanotechnology and for precision measurements of the Casimir force. The basic method for the calculation of the interaction forces is the Lifshitz formula \cite{Lif56}. It holds for planar or stratified material bodies having permittivity. Permeability  as well as frequency dispersion and finite temperature  may be included too. The generalization to non planar geometry is more complicated in case the variables in the underlying wave equation do not separate and for the interaction of two material bodies one is bound to the case of plane parallel geometry. However, most force measurements are done with a sphere or lens in front of a plane and there is a demand for generalizations to non planar geometry.

The most frequently used method for  non planar geometry is the proximity force approximation (PFA). Here the Casimir or van der Waals force known from planar geometry is taken at the local distance   and then averaged over  the surfaces. In this method the curvature of the surfaces in included in the sense of a first correction. The method holds obviously for small deviation from the plane parallel geometry. Its precision cannot be estimated since it is impossible to calculate higher order corrections by this method. 

The method of PFA dates back to Derjaguin \cite{derjaguin} and only recently it became possible to go beyond analytically. In \cite{Bordag:2006vc} the first correction beyond PFA was calculated for a cylinder in front of a plane with conductor boundary conditions. For the part containing Dirichlet boundary conditions this was confirmed numerically by the world line method \cite{GIES2006}. The opposite case of large and medium separations is easier and more detailed results are available, see \cite{WIRZBA2006} and \cite{EMIG2006}.

Another line of generalizations of the Lifshitz formula is to consider the interaction between two graphene sheets  \cite{Bordag:2005by} having in mind applications to carbon nanotubes. The graphene sheets are described by the two dimensional plasma shell model used in \cite{Barton:1991pb,BV}. For thin shells this model is more appropriate because the Casimir or van der Waals forces vanish if the thickness of a dielectric layer goes to zero. In \cite{Bordag:2005by}  the interaction between two such sheets was studied and in \cite{BORDAG2006D} this was generalized to the interaction of a material body with a flat graphene sheet and, further, to a cylindrical graphene sheet in front of  a plane in PFA.

In the present paper we derive the generalized Lifshitz formula in the first approximation beyond PFA for a cylindrical graphene sheet in front of a flat graphene sheet or a plane  material body with permittivity $\ep(\om)$. In both cases we consider two scalar problems corresponding to the TE and TM modes. Regrettably, in this configuration the polarizations of the electromagnetic field do not separate (in opposite to the case of a conducting wave guide). However for small separations the TM mode dominates and for large separations  the ideal conductor case is recovered.

The paper is organized as follows. In the next section we collect the basic formulas of the method of \cite{Bordag:2006vc} and generalize them to semitransparent boundaries. In section 3 we calculate the interaction energies in first approximation beyond PFA. After the conclusions some appendixes follow with details of the calculations.

\section{Functional integration and semitransparent boun\-daries}
In this section we start from the method of introducing boundary conditions into a functional integral by functional delta functions. The method was used in \cite{Bordag:1985zk} for the calculation of radiative corrections to the Casimir effect. Later it was rediscovered \cite{LI1992}. In \cite{EMIG2006} and \cite{Bordag:2006vc} it was used to  obtain a finite expression for the Casimir interaction energy of a sphere or a cylinder with a plane for Dirichlet and conductor boundary conditions. In \cite{WIRZBA2006} similar results were achieved in a multiscattering approach. Here we follow the representation given in \cite{Bordag:2006vc} and generalize it to semitransparent boundaries. 

Before entering the formalism of functional integration we formulate the boundary conditions which are considered in this paper. There are two types, that on the surface of a dielectric body and that on an infinitely thin plasma shell. We mention that both are in fact rather matching conditions relating the fields on both sides of the surface. These boundaries are semitransparent in the sense that the corresponding reflection and transmission coefficients take values somewhere in the interval from zero to unity. Both turn into Dirichlet resp. Neumann boundary conditions in the limiting case of the frequency parameter going to infinity. 
\begin{enumerate}
   \item Plasma shell\\
 Here we consider an infinitely thin sheet filled with a charged fluid in an oppositely charged immobile neutralizing background. This is the two dimensionale limiting case of three dimensional plasma used for example in the theory of metals. 
It is aimed to describe the $\pi$-electrons of a graphene sheet or a carbon nano tube. For details see \cite{BV} and papers cited therein.  The interaction of such a sheet with the electromagnetic field can be reduced to matching conditions across the sheet. For a flat sheet the polarizations separate into TE and TM modes and the matching conditions read
\be\label{mcdelta}\begin{array}{rcrlcrlr}
\Phi_+  &-&  \Phi_-&=&0, \qquad \Phi'_+\ - \ \Phi'_-&=&2\Om \ \Phi, &\qquad(TE)\\
\Phi'_+ &-&   \Phi'_-&=&0, \qquad \Phi_+\ - \ \Phi_-&=&-2\frac{\Om}{\om^2}\ \Phi', &\qquad(TM)
 \end{array}\ee
where
\be\label{Om}\Om=2\pi\frac{ne^2}{m}
\ee
is a parameter in parallel to the plasma frequency $\om_p$ in the second model. It depends on the parameters of the plasma, the density $n$ of the electrons, their charge $e$ and mass $m$. The inverse, $1/\Om$, is a length which is to be compared with the geometric sizes of the interacting bodies. For a carbon nano tube we note $1/\Om= 1.5\mu m$ (see \cite{BORDAG2006D}, Eq.(11)), which is by two orders of magnitude lager that the corresponding value, say  $1/\om_p=0.02\mu m$ for gold, in the second model. 
%
The matching conditions for the TE polarization is just the same as for a delta function potential of strength $\Om$ for a scalar field 
and up to some subtleties  the conditions for the TM mode correspond to a potential with the derivative of a delta function. As shortcut we will use the subscripts $\delta_{\rm TE}$ and $\delta_{\rm TM}$. 

In addition to a flat sheet we consider also a cylindrical one.  In that case the polarizations do not separate (in opposite to a conducting wave guide).  We define the analogous scalar problems by the same matching conditions \Ref{mcdelta} as in the case of a flat boundary  on the radial functions,
\be\label{Fcyl}\Phi_{\om}(x,y,z)=\int\frac{dk_{z}}{2\pi}\sum_{m=-\infty}^{\infty} \ e^{i k_{z} z +i m \varphi} \ \Phi_{\om,k_{z},m}(r)
\ee
with polar coordinates $(r,\varphi)$ in the plane perpendicular the the axis of the cylinder, $x_1=r\cos\varphi$, $x_2=r\sin\varphi$. The prime denotes then the radial derivative.
 \item Dielectric body\\
  We consider a dielectric body with permittivity $\ep$ filling the half space $x_1>0$,
\be\label{epz}\ep(x_1,\om)=\left\{ {\ep(\om)\quad\mbox{for} \ x_1>0,\atop 0\quad\mbox{~~~ for} \  x_1<0.}\right.
\ee
In order to consider later the short separation limit we need to include frequency dispersion.  We take the plasma model,
\be\label{plm}\ep(\om)=1-\frac{\om_p^2}{\om^2},
\ee
where $\om_p$ is the plasma frequency. Usually, this model is used for metals.

Since the surface is flat the polarizations of the electromagnetic field separate into transverse electric (TE) and transverse magnetic (TM) modes. The well known matching conditions  for the corresponding scalar functions (amplitudes) $\Phi(z)$ are
\be\label{mcep}\begin{array}{rcrlcrlr}
\Phi_+  &-&  \Phi_-&=&0, \qquad \Phi'_+\ - \ \Phi'_-&=&0, &\qquad(TE)\\
\ep \Phi_+ &-&   \Phi_-&=&0, \qquad \Phi'_+\ - \ \Phi'_-&=&0, &\qquad(TM)
 \end{array}\ee
where $\Phi_\pm$  are the limiting values from the right and from the left of the plane   and the prime denotes the derivative with respect to $x_1$. Here the Fourier transform in the translational invariant directions $x_{||}=(x_2,x_3)$, i.e., in the directions parallel to the plane, is assumed,
\be\label{Fflat}\Phi_{\om}(x_1,x_2,x_3)=\int\frac{dk_{||}}{2\pi} \ e^{i k_{||} x_{||}} \ \Phi_{\om,k_{||}}(x_1).
\ee
The amplitudes satisfy the wave equation
\be\label{eom}\left( \ep(z,\om)\ \om^2-k_{||}^2+\frac{\pa^2}{\pa x_1^2}\right) \Phi_{\om,k_{||}}(x_1)=0.
\ee
In the following we use the shortcuts $\epsilon_{\rm TE}$ and $\ep_{\rm TM}$ to identify these models.
\end{enumerate}

The method of implementing boundary or matching conditions in the functional integration rests on the restriction of the integration space in a functional integral%
\be
\label{Z1}Z(J)=\int D\phi \ \prod_{ S}\delta\left(H_z[\phi]\right) \ exp\{- \S \}
\ee
representing the generating functional $Z$ (or the partition function) of a field theory with a field $\Phi$ and some action $\S$. Initially, the theory is completely arbitrary and it may contain interaction, background fields and boundaries. The functional delta function provides then the necessary restriction of the integration space to fields fulfilling additional  boundary or matching conditions on a surface $S$. We assume that these conditions can be formulated in terms of a linear functional,
\be\label{Hphi}H_z[\Phi]\equiv\int H(z,x)\Phi(x) \ dx=0 \ ,  
\ee
where $H(z,x)$ is some integral kernel. For example, Dirichlet boundary conditions are given by
\be\label{}H^{\rm D}(z,x)=\delta(x-f(z)),
\ee
where the function $f(z)$ describes the surface $S=\left\{x| \ x=f(z)\right\}$  and $z$ provides a parameterization of the surface $S$. Assuming the surface $S$ is a plane parallel to the $(x_2,x_3)$-plane at $x_1=a$ in an $\mathbb{R}^3$.  For that a parameterization is $z=\left(\begin{array}{c}x_2\\x_3\end{array}\right)$ and $f(z)=\left(\begin{array}{c}a\\x_2\\x_3\end{array}\right)$. In fact, below we need the corresponding quantities for a cylinder of radius $R$ whose axis coincides with the $z$-axis: $z=(\varphi,x_3)$ and $f(z)=\left(\begin{array}{c}R\cos\varphi\\R\sin\varphi\\x_3\end{array}\right)$.

The function $H(z,x)$ for the upper line of the matching conditions \Ref{mcdelta} reads
\be\label{HTE}H^{\delta_{\rm TE}}(z,x)=
\delta\left(x-f(z)\right)\left(\pa_{n_+}-\pa_{n_-}-2\Om \right),
\ee
where $\pa_{n_+}$ is the normal derivative on the right side of the surface ($x_1>a$) and $\pa_{n_-}$ is that on the left side.
The function  $H(z,x)$ for the lower line of the matching conditions \Ref{mcdelta} reads
\be\label{HTM}H^{\delta_{\rm TM}}(z,x)=
\delta\left(x-f(z)_+\right)-\delta\left(x-f(z)_-\right)
+\frac{2\Om}{\om^2}\ \delta\left(x-f(z)\right)\pa_n.
\ee
Here $f(z)_+$ assumes to take the limiting value if $x$ approaches the surface from the right side. The normal derivative can be taken on either side. Using these functions in Eq. \Ref{Hphi} one obtains just the matching conditions \Ref{mcdelta} for a plane or, with the appropriate choice of $z$ and $f(z)$, for a cylinder. 

It should be mentioned that this method is completely general and that it holds for any boundary or matching conditions which can be written in form of a linear functional like \Ref{Hphi}. Also, it is not restricted to a scalar field. Originally it was used in  \cite{Bordag:1985zk} for the electromagnetic field.

However, there is also a limitation of this method. It cannot been used for the matching conditions \Ref{mcep} on the surface of a dielectric body since there are different speeds of light on both sides and the condition cannot be  expressed in form of a delta function in the functional integral. Whether this limitation can be overcome is not known at the moment. There exists an attempt to incorporate a dielectric body into the functional integral \cite{EMIG2004} but it looks too complicated and no nontrivial example was given so far. Also the recent approach of \cite{KENNETH2006} cannot solve this problem since it cannot handle the TM mode.

With the formulation \Ref{Hphi} of the boundary conditions the method follows exactly the steps given in section 2 of \cite{Bordag:2006vc}. One arrives at a representation of the Casimir energy in the form
\be\label{Ecas}E_{\rm Cas}=\frac12 \int_{-\infty}^\infty\frac{d\om}{2\pi} \ tr_S \ln K,
\ee
where $K$ has an integral kernel,
\be\label{K}K(z,z')=\int dx \ dy \ \ H(z,x)D(x,y)H^\top (y,z'),
 \ee
in the space of functions defined on $S$ and the trace is over such functions. $D(x,y)$ is the propagator of the initial theory defined by the action $\S$ in \Ref{Z1} and the transposition in $H^\top (y,z')$ means that the derivatives in $H$ act to the left.
A difference of this formula as compared to Eq.(15) in \cite{Bordag:2006vc} is that we already passed from time dependency to frequency dependent quantities by means of the corresponding Fourier transform. Also, starting from here we work in the Euclidean version.

\begin{figure}\unitlength1cm
 \begin{picture}(14,5)
 \put(0,0){ \includegraphics[width=7cm]{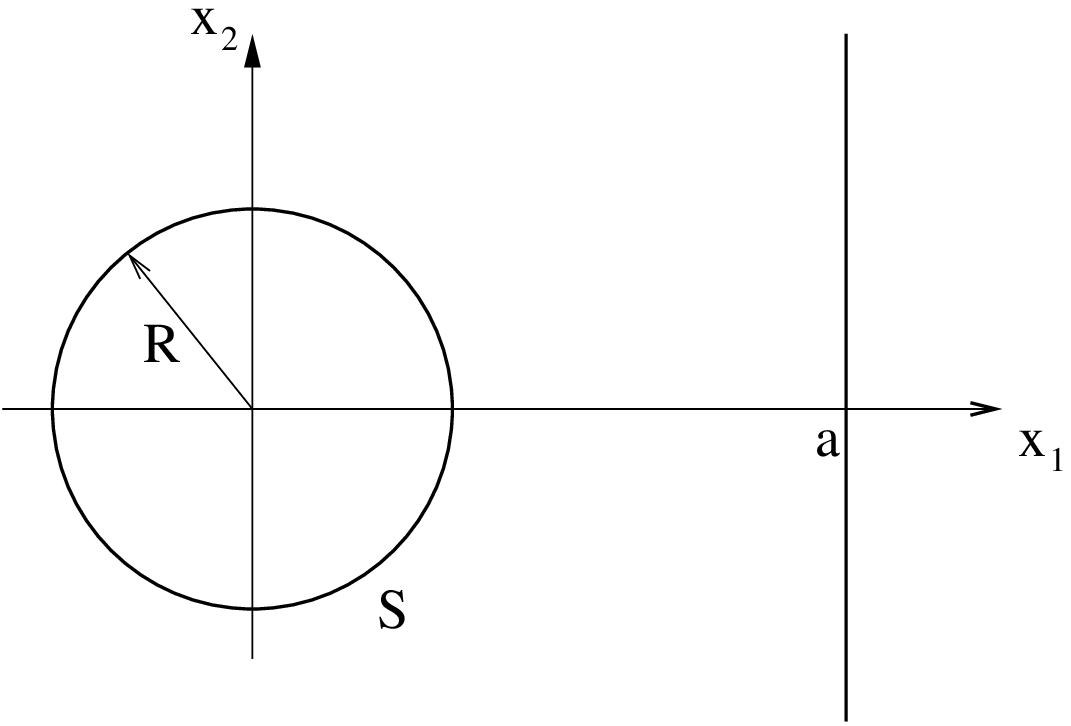}}
 \end{picture}
 \caption{The configuration of a cylinder in front of the plane $S$} \label{fig0}
\end{figure}

The configuration we are interested in is shown in Fig. \ref{fig0} and we proceed as follows. We consider  a free scalar field in the presence of a dielectric halfspace at $x_1>a$   or a flat plasma shell at $x_1=a$ given by the action $\S$ and the functional integral \Ref{Z1}  without the functional delta function as the initial theory.
Further we consider this initial theory to interact with a cylindrical plasma shell  $S$ with radius $R$. This interaction is incorporated in \Ref{Z1} as matching conditions \Ref{mcdelta} using the functional delta function. In this way we formulated two problems, the interaction between a cylindrical plasma sheet $S$ and
\begin{enumerate}
  \item a flat plasma sheet
  (shortcuts $(\delta\delta)_{\rm TE}$ and  $(\delta\delta)_{\rm TM}$),
 \item a dielectric halfspace 
  (shortcuts $(\ep\delta)_{\rm TE}$ and  $(\ep\delta)_{\rm TM}$).
 \end{enumerate}
In the following, to a large extend the formulas are the same for all considered problems. 
The shortcuts will be used for the quantities specific for the considered problem. 

In general, other combinations, for example the interaction of a cylindrical shell obeying TE conditions with a plane carrying a TM condition, can be considered too but we restrict ourselves here to the cases which appear in the physical applications. 

To continue we have to specify the propagator $D(x,x')$ of the initial theory, i.e., for a flat plasma shell and for a dielectric halfspace. These are well known expressions but we need them is a somewhat specific representation, namely as a difference
\be\label{D}D_{\om}(x,x')=D^{(0)}_{\om}(x-x')-\tilde{D}_{\om}(x,x'),
\ee
where the subscript $\om$ indicates that the Fourier transform in the time variable was done.
The first term in the r.h.s., $D^{(0)}(x-x')$, is the free space propagator for a massless scalar field,
\be\label{D0}D^{(0)}(x-x')=\int\frac{d^3k}{(2\pi)^3} \ \frac{e^{-ik(x-x')}}{\om^2+k^2}.
\ee
In the following we need the representations
\be\label{D0f}D^{(0)}_{\om}(x-x')=\int\frac{d^2k_{||}}{(2\pi)^2} \ \ e^{-ik_{||}(x_{||}-x'_{||})} \ d^{(0)}_{\om,k_{||}}(x_1-x_1')
\ee
with
\be\label{d0f}d^{(0)}_{\om,k_{||}}(x_1-x_1')=\frac{e^{-\gamma|x_1-x_1'|}}{2\gamma}
\qquad \left(\gamma=\sqrt{\om^2+k_{||}^2} \ \right)\ ,
\ee
which emerges from \Ref{D0} after integration over $k_1$ and
%
\be\label{D0c}D^{(0)}_{\om}(x-x')=\int\frac{dk_{3}}{(2\pi)^2} \ \ \sum_{m=-\infty}^\infty \ e^{-ik_{3}(x_{3}-x'_{3})+im(\varphi-\varphi')} \ d^{(0)}_{\om,k_{3},m}(r,r')
\ee
with
\be\label{d0c}d^{(0)}_{\om,k_{3},m}(r,r')=
\left\{{I_m(\rho r)K_m(\rho r') \quad \mbox{for}  \quad r'>r \ ,  \atop
I_m(\rho r')K_m(\rho r) \quad \mbox{for}\quad  r>r' \ , }  \right. 
\qquad \left(\rho=\sqrt{\om^2+k_{3}^2}\right) \ ,
\ee
which is the representation of the propagator in cylindrical coordinates in terms of modified Bessel functions. 

In Eq.\Ref{D} the addendum $\tilde{D}_{\om}(x,x')$ describes the boundary dependence of the propagator. For the flat plasma sheet we make use of the translational invariance in the directions parallel to the sheet and write down a representation in parallel to \Ref{D0f}, 
\be\label{Dti}\tilde{D}_{\om}(x,x')=
\int\frac{dk_{||}}{(2\pi)^2} \ \ e^{-ik_{||}(x_{||}-x'_{||})} \ \tilde{d}_{\om,k_{||}}(x_1,x_1')
\ee
and similar for $D_{\om}(x,x')$. Then the difference
\be\label{23}  d_{\om,k_{||}}(x_1,x_1')=
d^{(0)}_{\om,k_{||}}(x_1-x_1')-\tilde{d}_{\om,k_{||}}(x_1,x_1')
\ee
is subject to the same boundary or matching conditions as before.

For simplicity we start from Dirichlet boundary conditions. In that case the representation
\be\label{dD}
\tilde{d}^{\rm (Dir)}_{\om,k_{||}}(x_1,x_1')=
\frac{d^{(0)}_{\om,k_{||}}(x_1-a) \ d^{(0)}_{\om,k_{||}}(a-x_1')}{
d^{(0)}_{\om,k_{||}}(0)}
\ee
holds which can be checked by applying the boundary conditions, 
i.e., \Ref{23} with \Ref{dD} inserted  vanishes if either $x_1=a$ or $x_1'=a$ holds.  Note that this representation holds for the arguments $x_1$ and $x_1'$ on any side of the plane at $x_1=a$. For both arguments to the left of the plane the expression simplifies,
\be\label{dD1}\tilde{d}^{\rm (Dir)}_{\om,k_{||}}(x_1,x_1')=
\frac{1}{2\ga} \ e^{-\ga(2a-x_1-x_1')}
 \qquad (x_1,x_1'<a) \ .
\ee
Inserted together with \Ref{d0f} into \Ref{23} and, further, into \Ref{D0f}, just the well known propagator with one reflection on the mirror at $x_1=a$ appears.
Now we write down a similar expression for the  the plasma shell. For the TE matching conditions (first line in \Ref{mcdelta}) we obtain
\be\label{dd}\tilde{d}^{\, \delta_{\rm TE}}_{\om,k_{||}}(x_1,x_1')=
\frac{1}{2\ga} \ \frac{1}{1+\ga/\Om}
\ e^{-\ga\left(\mid x_1-a\mid+\mid x_1'-a\mid\right)},
\ee
which can be checked again by applying the matching condition. 
For both arguments to the left of the plane the expression simplifies,
\be\label{dtdeTE}\tilde{d}^{\, \delta_{\rm TE}}_{\om,k_{||}}(x_1,x_1')=
\frac{1}{2\ga} \ \frac{1}{1+\ga/\Om}
\ e^{-\ga(2a-x_1-x_1')}
 \qquad (x_1,x_1'<a) 
 \ .
\ee
Similar expressions can be found in \cite{Bordag:1992cm} where the propagator for a delta function potential was considered.

For the TM matching condition (second line in \Ref{mcdelta}) we proceed in the same way and obtain
\be\label{}\tilde{d}^{ (\delta_{\rm TM})}_{\om,k_{||}}(x_1,x_1')=
\frac{1}{2\ga} \ \frac{-\ga\Om}{\om^2+\ga \Om} \
\mbox{sign}(x_1-a) \ \mbox{sign}(x_1'-a) \
\ e^{-\ga\left(\mid x_1-a\mid+\mid x_1'-a\mid\right)},
\ee
which can also be checked  by applying the matching condition (note that we are here in the Euclidean version and $\om^2$ enters with the apposite sign as compared to Eq.\Ref{mcdelta}).  We note that $\tilde{d}^{ (\delta_{\rm TM})}_{\om,k_{||}}(x_1,x_1')$ has a jump if $x_1$ passes through $a$ whereas the derivative is continuous. To the left of the plane the representation simplifies too,
\be\label{dtdeTM}\tilde{d}^{ (\delta_{\rm TM})}_{\om,k_{||}}(x_1,x_1')=
\frac{1}{2\ga} \ \frac{-\ga\Om}{\om^2+\ga \Om} \
\ e^{-\ga\left(2a-x_1-x_1' \right)}  \qquad (x_1,x_1'<a)  \ .
\ee
Now we need the corresponding formulas for the dielectric half space. These are standard and can be found in many places. For the TE mode it holds
\be\label{dtepTE}\tilde{d}^{\, \ep_{\rm TE}}_{\om,k_{||}}(x_1,x_1')=
\frac{1}{2\ga} \ \frac{\ga-p}{\ga+p} \
\ e^{-\ga\left(2a-x_1-x_1'\right)}  \qquad (x_1,x_1'<a)  \ 
\ee
with $p=\sqrt{\ep(i\om)\om^2+k_{||}^2}$ which is the momentum perpendicular to the plane in the medium after rotation to the imaginary axis. For the TM mode we note
\be\label{dtepTM}\tilde{d}^{\, \ep_{\rm TM}}_{\om,k_{||}}(x_1,x_1')=
\frac{-1}{2\ga} \ \frac{\ep(i\om)\ga-p}{\ep(i\om)\ga+p} \
\ e^{-\ga\left(2a-x_1-x_1'\right)}  \qquad (x_1,x_1'<a)  \ .
\ee
Obviously, the formulas \Ref{dtdeTE}, \Ref{dtdeTM}, \Ref{dtepTE} and \Ref{dtepTM} have the same dependence on $x_1$ and $x_1'$ and they can be joined into
\be\label{dtallg}
\tilde{d}_{\om,k_{||}}(x_1,x_1')=
\tilde{d}_{\om,\ga} \ 
\ e^{-\ga\left(2a-x_1-x_1'\right)}  \qquad (x_1,x_1'<a) 
\ee
with $\tilde{d}_{\om,\ga}$ to be substituted by one of
\be\label{dfour}
\begin{array}{rclrcl}
\tilde{d}^{\, \delta_{\rm TE}}_{\om,\ga}&=&\frac{1}{2\ga} \ \frac{1}{1+\ga/\Om} \ , & \qquad
\tilde{d}^{\, \delta_{\rm TM}}_{\om,\ga}&=&\frac{1}{2\ga} \ \frac{-\ga\Om}{\om^2+\ga \Om} \ , \\ [8pt]
\tilde{d}^{\, \ep_{\rm TE}}_{\om,\ga}&=&\frac{1}{2\ga} \ \frac{\ga-p}{\ga+p} \ , &
\tilde{d}^{\, \ep_{\rm TM}}_{\om,\ga}&=&\frac{-1}{2\ga} \ \frac{\ep(i\om)\ga-p}{\ep(i\om)\ga+p} \ .
\end{array}
\ee
Up to a common factor these are just the reflection coefficients of the related scattering problems. 

Now with the explizite expressions for the propagator $D(x,x')$ of the initial theory at hand we return to Eq.\Ref{Ecas}. The next step is to calculate the trace over $K(z,z')$. We chose a convenient basis for that, namely
\be\label{k3m}\mid k_3,m\rangle=\frac{1}{2\pi} \ e^{ik_3x_3+i m \varphi}
\ee
with respect to which all functions defined on the cylinder $S$ can be expanded. For details see \cite{EMIG2006,Bordag:2006vc}. Then Eq. \Ref{Ecas} can be written in the form
\be\label{Ecas1a}E_{\rm Cas}=\frac12\int\frac{d\om}{2\pi}\int\frac{dk_3}{2\pi} \ 
tr_m\ln K_{m,m'}
\ee
with
\be\label{}K_{m,m'}=\langle k_3,m\mid K(z,z')\mid k_3,m'\rangle \ .
\ee
We mention that $K_{m,m'}$ is an infinite dimensional matrix labeled by $m$ and $m'$ and that $\ln K_{m,m'}$ is another matrix and that the trace is over the latter matrix. In Eq.\Ref{Ecas1a} the translational invariance along the axis of the cylinder was taken into account and $E_{\rm Cas}$,\Ref{Ecas1a}, is in fact the energy density per unit length of the cylinder. 

We proceed with dividing $K_{m,m'}$ into two parts according to the subdivision in Eq.\Ref{D} and \Ref{K},
\be\label{K1}K_{m,m'}=K^{(0)}_{m,m'}-\tilde{K}_{m,m'}
\ee
with
\bea\label{K2}
K^{(0)}_{m,m'}&=&\langle k_3,m\mid K^{(0)}(z,z')\mid k_3,m'\rangle ,\nn \\
\tilde{K}_{m,m'}&=&\langle k_3,m\mid \tilde{K}(z,z')\mid k_3,m'\rangle .
\eea
Now we remark that the functions $H(z,x)$ appearing in the formulation \Ref{Hphi} of the matching conditions commute with the averaging in the basis \Ref{k3m} and we obtain
\be\label{}K_{m,m'}=\int dr \ dr' \ H(r) \  d_{\om,k_3,m,m'}(r,r') \ H^\top(r')
\ee
with
\be\label{}d_{\om,k_3,m,m'}(r,r')=\langle k_3,m\mid D_\om(x,x')\mid k_3,m'\rangle .
\ee
With the same subdivision as in Eq.\Ref{D} into free space part and addendum depending on the boundary conditions on the plane in $x_1=a$ we define
\be\label{}d_{\om,k_3,m,m'}(r,r')=
d^{(0)}_{\om,k_3,m}(r,r')\delta_{m,m'}-\tilde{d}_{\om,k_3,m,m'}(r,r'),
\ee
where $d^{(0)}_{\om,k_3,m}(r,r')$ is the free space part introduced in Eq.\Ref{D0c}. Obviously its contribution is diagonal in $m$ and $m'$. 

Next we need the functions $H(z,x)$, Eqs.\Ref{HTE} and \Ref{HTM} for the cylindrical surface $S$. They read
\bea\label{Hrad}H^{\delta_{\rm TE}}(r)&=&
\delta(r-R)\left(\pa_{r_+}-\pa_{r_-}-2\Om\right),
\nn \\ 
H^{\delta_{\rm TM}}(r)&=&
\delta(r-(R+0))-\delta(r-(R-0))-\frac{2\Om}{\om^2}\delta(r-(R+0)).
\eea
Now we calculate $K^{(0)}_{m,m'}$. For the TE case we have
\be\label{}K^{(0){\rm (TE)}}_{m,m'}=\delta_{m,m'}K^{(0){\rm (TE)}}_{m}
\ee
with
\be
K^{(0){\rm (TE)}}_{m}=
\int dr \ dr' \ \ 
 H^{\delta_{\rm TE}}(r) \ d^{(0)}_{\om,k_3,m}(r,r') \ {H^{\delta_{\rm TE}}}^\top(r').
\ee
First we remark that in applying ${H^{\delta_{\rm TE}}}^\top(r')$ we have to keep $r\ne R$ so that $d^{(0)}_{\om,k_3,m}(r,r')$ is continuous at $r=R$ including its derivatives and we come to
\be\label{}K^{(0){\rm (TE)}}_{m}=\int dr \ \ 
 H^{\delta_{\rm TE}}(r) \ d^{(0)}_{\om,k_3,m}(r,R) \ (-2\Om).
\ee
Now we apply $H^{\delta_{\rm TE}}(r)$ and we have to take care of the jump of the derivative with respect to $r$ of $d^{(0)}_{\om,k_3,m}(r,R)$, see Eq.\Ref{d0c}. Using $I_m'(z)K_m(z)-I_m(z)K_m'(z)=1/z$ we get
\be\label{K0TE}
K^{(0){\rm (TE)}}_{m}=\frac{2\Om}{R}\left(1+2\Om R \ I_m(\rho R) \ K_m(\rho R)\right).
\ee
Proceeding in the same way with the TM mode  first we  define
\be\label{}K^{(0){\rm (TM)}}_{m,m'}=\delta_{m,m'}K^{(0){\rm (TM)}}_{m}
\ee
with
\be
K^{(0){\rm (TM)}}_{m}=
\int dr \ dr' \ \ 
 H^{\delta_{\rm TM}}(r) \ d^{(0)}_{\om,k_3,m}(r,r') \ {H^{\delta_{\rm TM}}}^\top(r').
\ee
Application of $ {H^{\delta_{\rm TM}}}^\top(r')$ gives 
\be\label{}K^{(0){\rm (TM)}}_{m}= \int dr \ \ 
 H^{\delta_{\rm TM}}(r) \ d^{(0)}_{\om,k_3,m}(r,r') \ 
{\raisebox{4pt}{$\begin{array}{c}\leftarrow\\ [-6pt] \pa\end{array}$}_{\! \! \! \! \! \! r'}}_{ {\Big |_{r'=R}  }} \ 
\frac{2\Om}{-\om^2}
\ee
and finally we obtain
\be\label{K0TM}K^{(0){\rm (TM)}}_{m}=
\frac{-2\Om}{\om^2 R}
\left(1-\frac{2\Om \rho^2 R}{\om^2} \ I_m'(\rho R)K_m'(\rho R)\right).
\ee
In this way we calculated the projection of the free space part of the propagator onto the cylinder with the matching conditions of the plamsa sheet. These expressions alone, i.e., $K^{(0){\rm (TE)}}_{m}$ or $K^{(0){\rm (TM)}}_{m}$, inserted into Eq.\Ref{Ecas1a} would give the Casimir energy of a cylindrical plasma shell (more extactly, since the polarizations of the electromagnetic field do not separate, the Casimir energies of the two scalar problems defined by the matching conditions \Ref{mcdelta}). This Casimir energy contains ultraviolet divergences which, b.t.w., are not investigated yet.

Now we calculate the second part in \Ref{K1} which contains the information on the plane in $x_1=a$. Here, of course, we do not have the cylindrical symmetry. For the propagator $\tilde{D}(x,x')$ in $\tilde{K}_{m,m'}$, Eqs.\Ref{K2}, \Ref{K}, we use the representations \Ref{Dti} with \Ref{dtallg}. We note that   $\tilde{D}(x,x')$  and its derivatives are continuous at the cylinder (the jumps are on the plane at $x_1=a$). Therefore, in each function $H(r)$, Eq.\Ref{Hrad}, only the last term contributes. In this way we come to 
\be\label{HonTE}
H^{\delta_{\rm TE}}\tilde{D}(x,x') {H^{\delta_{\rm TE}}}^\top=
\left(-2\Om\right)^2 \tilde{D}(x,x')_{\big |_{r=r'=R}}
\ee
and 
\be\label{HonTM}
H^{\delta_{\rm TM}}\tilde{D}(x,x') {H^{\delta_{\rm TM}}}^\top=
\left(\frac{2\Om}{\om^2}\right)^2 
\pa_r \pa_{r'}\tilde{D}(x,x')_{\big |_{r=r'=R}} \ 
\ee
with the already mentioned cylindrical coordinates $x_1=r\cos\varphi$, $x_2=r\sin\varphi$, $x_1'=r'\cos\varphi'$, $x_2'=r'\sin\varphi'$.

We have to take these expression in the basis \Ref{k3m} and obtain with \Ref{Dti} and \Ref{dtallg}
\bea\label{}\tilde{K}^{\rm (TE)}_{m,m'}&=&
\langle k_3,m\mid (-2\Om)^2 
\tilde{D}(x,x')_{\big |_{r=r'=R}}
\mid k_3,m'\rangle
\nn \\ &=&
4\Om^2\int dk_2 \ f_m f^*_{m'} \ e^{-2\ga a} \tilde{d}_{\om,\ga}
\eea
with
\be\label{}f_m=\int_0^{2\pi} \frac{d\varphi}{2\pi} \ e^{-im\varphi+ik_2x_2+\ga x_1}.
\ee
Introducing new variables
\bea\label{theta}
k_2&=&\rho \sinh \theta ,\nn \\
\ga&=&\rho \cosh \theta ,
\eea
the transforming the integration over $\varphi$ into an integral representation of the modified Bessel function, the $f_m$ become
\be\label{fm}f_m=e^{m \theta}I_m(\rho r).
\ee
With this the final representation of $\tilde{K}^{\rm (TE)}_{m,m'}$ is
\be\label{KtTE}\tilde{K}^{\rm (TE)}_{m,m'}=4\Om^2 \ I_m(\rho R) \ I_{m'}(\rho R)\ \A_{m+m'}
\ee
with
\be\label{AT}\A_{m+m'}=\int_0^\infty d\theta \ \cosh\left((m+m')\theta\right)
\ 2\rho\cosh\theta \ \tilde{d}_{\om,\ga} \ e^{-2a\rho\cosh\theta}.
\ee
We note that in the limiting case of $\Om\to\infty$ where the matching conditions \Ref{mcdelta} turn into Dirichlet boundary conditions, $2\rho\cosh\theta \ \tilde{d}_{\om,\ga}\to 1$ holds and $\A_{m+m'}$ \Ref{AT} becomes an integral representation of the modified Bessel function,
\be\label{Kh}\A_{m+m'} \raisebox{-4pt}{${\longrightarrow\atop\Om\to\infty}$}K_{m+m'}(2a\rho),
\ee
which coincides with the corresponding formulas in \cite{EMIG2006} and \cite{Bordag:2006vc}. However, for finite $\Om$ we are left with the integral representation \Ref{AT}. 

For the TM case we proceed in the same way. After applying the functions $H(r)$ \Ref{HonTM} and projecting on the basis $\mid k_3,m\rangle$ we get
\be\label{}\tilde{K}^{\rm (TM)}_{m,m'}=
\left(\frac{2\Om}{\om^2}\right)^2 \pa_r \pa_{r'} 
\langle k_3,m\mid 
\tilde{D}(x,x')_{\big |_{r=r'=R}}
\mid k_3,m'\rangle \ .
\ee
The matrix elements can be calculated in parallel to the TE case. The radial derivatives apply only to the Bessel functions $I_m(\rho r)$ since $\A_{m+m'}$ does not carry any dependence on $r$ and we end up with
\be\label{KtTM}\tilde{K}^{\rm (TM)}_{m,m'}=
\frac{4\Om^2\rho^2}{\om^4} \ I_m'(\rho R)I_{m'}'(\rho R) \ \A_{m+m'} .
\ee
Again, in the limiting case $\Om\to\infty$ we reobtain the corresponding expression for hard boundary conditions, Neumann  ones in this case. 

By means of formulas \Ref{K0TE}, \Ref{K0TM} and \Ref{KtTE},\Ref{KtTM} we calculated the matrixes entering $K_{m,m'}$ in Eq.\Ref{K1}. Since $K^{(0)}_{m,m'}$ is diagonal we can easily separate it and the logarithm in Eq.\Ref{Ecas1a} becomes
\be\label{sepln}\ln K_{m,m'}= 
\ln \left(K^{(0)}_{m}\delta_{m,m'}\right)
+\ln\left(\delta_{m,m'}-A_{m,m'}\right)
\ee
with 
\be\label{}A_{m,m'}=\frac{1}{K^{(0)}_{m}}\ \tilde{K}_{m,m'}.
\ee
In \Ref{sepln} the first term gives the Casimir energy of the plasma shell cylinder alone. The decisive point of the method is now that this part carries all ultraviolet divergences which are in $E_{\rm Cas}$, Eq.\Ref{Ecas}, and that it does not depend on the distance between the plane and the cylinder. Therefore the distance dependence is solely  contained in  the second term in the r.h.s. of \Ref{sepln}. It gives rise to  the distance dependent part of the Casimir energy, 
\be\label{Ecas1}
E_{\rm Cas}=\frac12\int\frac{d\om}{2\pi}\int\frac{dk_3}{2\pi} \ 
tr_m\ln \left(\delta_{m,m'}-A_{m,m'}\right),
\ee
which  does not contain any ultraviolet divergence. It serves as the basic representation for the further elaboration.

Inserting \Ref{K0TE},  \Ref{K0TM} and \Ref{KtTE},\Ref{KtTM} we simplify the expressions to some extend,
\be\label{ATE}A_{m,m'}^{\delta_{\rm TE}}=R^{\delta_{\rm TE}} \ \ \A_{m+m'}
\ee
with
\be\label{}R^{\delta_{\rm TE}}=\frac{1}{1+\frac{1}{2\Om R I_m(\rho R)K_{m}(\rho R)}} \ \frac{I_{m'}(\rho R)}{K_{m}(\rho R)} 
\ee
and
\be\label{ATM}A_{m,m'}^{\delta_{\rm TM}}= R^{\delta_{\rm TM}}\ \A_{m+m'}
\ee
with
\be\label{}R^{\delta_{\rm TM}}=
\frac{1}{1-\frac{\om^2}{2\Om R \rho^2}\frac{1}{ I_m'(\rho R)K_{m}'(\rho R)}} \ \frac{I_{m'}'(\rho R)}{K_{m}'(\rho R)}.
\ee
In this representation the information on the cylindrical plasma shell is in the first factors on the r.h.s., i.e., in $R^{\delta_{\rm TE}}$ and in $R^{\delta_{\rm TM}}$,  whereas the $ \A_{m+m'}$ by means of Eqs.\Ref{AT} and \Ref{dtallg} carry the information on the plane at $x_1=a$, whose different cases will be indicated in the following by a corresponding superscript, 
$ \A_{m+m'}^{\delta_{\rm TE}}$ or $ \A_{m+m'}^{\ep_{\rm TE}}$ for example. 

\section{Small separation and first correction beyond PFA}
In this section we generalize the small separation expansion of section IV in  \meins ~to the semitransparent boundaries of a plasma sheet and a dielectric half space interacting with a cylindrical plasma sheet. For the dielectric halfspace with dispersion $\ep(\om)$ we restrict ourselves now to the plasma model, i.e., to a permittivity given by Eq.\Ref{plm}. We start from the representation \Ref{Ecas1} of the interaction Casimir energy with the matrix elements $A_{m,m'}^{\delta_{\rm TE}}$ and  $A_{m,m'}^{\delta_{\rm TM}}$  given by Eqs. \Ref{ATE} and \Ref{ATM}. The idea of the small separation expansion is is that only high momenta, i.e., large all, $\om$, $k_3$, $m$ and $m'$, contribute and that one can use the asymptotic expansions of the matrix elements. Further, it is useful to expand the logarithm although at the end it needs to be summed up again. We note that this asymptotic expansion is the same as done for hard boundary conditions in \meins, however the expressions depend in addition on the parameters $\Om$ and $\om_p$ and they are more involved.

We start by expanding the logarithm in Eq.\Ref{Ecas1},
\be\label{Ecas2}E_{\rm Cas}=\frac{-1}{2}
\int_{-\infty}^\infty\frac{d\om}{2\pi}\int_{-\infty}^\infty\frac{dk_3}{2\pi}
\sum_{s=0}^\infty\frac{1}{s+1} \int_{-\infty}^\infty dm \ 
\int_{-\infty}^\infty dn_1 \ \dots \int_{-\infty}^\infty dn_s \ 
\M
\ee
with
\be\label{M}\M=A_{m,m+n_1}A_{m+n_1,m+n_2}\dots A_{m+n_{s-1},m+n_{s}}A_{m+n_{s},m} .
\ee
The next step is to make the rescaling $\om\to\om/R$ and $k_3\to k_3/R$ which makes $R$ disappear everywhere except for a factor $1/R^2$ in front of the whole expression  and for the exponential in $A_{m+m'}$, Eq.\Ref{AT}, where it combines with $a$ into
\be\label{}\frac{a}{R}=\frac{R+L}{R}= 1+\ep,
\ee
where we introduced the distance $L$  between the cylinder and the plane and $\ep=\frac{L}{R}$, which is the small parameter at short separation.
The other two length parameters in the considered problem are $1/\Om$ and $1/\om_p$. Of interest are values in the nanometer and micrometer region. The short separation expansion implies  $1/\Om<<R$ and $1/\om_p<<R$. However, with respect to the separation $L$ we assume them to be of the same order, i.,e., $1/\Om\sim L$ and  $1/\om_p\sim L$. As a consequence we have to consider the combinations $\Om R \ep=\Om L\equiv\Om_L$ and $\om_p R\ep=\om_p L\equiv\om_L$ which will appear in the following as being of order of one. The latter notations will be used in the intermediate steps. In order to shorten the notations, in the following we will put $R=1$. The correct dimension of the energy can be restored by 
$E_{\rm Cas}\to E_{\rm Cas}/R^2$.  In the final expressions, only the  parameters $\Om_L$ and $\om_L$ will appear and their dimensions are restored by $\Om_L\to \Om L$ and $\om_L\to \om_p L$. 

We proceed with the following substitutions of variables in \Ref{Ecas2}. Taking into account that the integrand is an even function of $\om$, $k_3$ and $m$  we first substitute
$\om=\rho\sin\al$ and $k_3=\rho\cos\al$
and then
\be\label{varsub}\rho=\frac{t}{\ep}\sqrt{1-\tau^2}, \ \ m=\frac{t}{\ep}\tau, \ \ \  n_i\to n_i\sqrt{\frac{4t}{\ep}}
\ee
after which the Casimir energy can be written in the form
\bea\label{ECas3}E_{\rm Cas}&=& 
\frac{-\ep^{-3}}{2\pi}\sum_{s=0}^\infty\frac{1}{s+1}
\int_0^\infty\frac{dt}{t}t^3
\int_0^{\pi/2}\frac{d\al}{\pi/2}\int_0^1d\tau \ 
\nn \\ && ~~~~~~
    \times
\int_{-\infty}^\infty n_1 \ \dots \int_{-\infty}^\infty n_s \
\left(\frac{4t}{\ep}\right)^{s/2} 
\M \ ,
\eea
where $\M$ is still given by Eq.\Ref{M}. We note that in the TE case $\M$ does not depend on $\al$ and $\tau$ and these integrations give a factor of unity. However, in the TM case $\M$ depends on both these variables. 

In Eq.\Ref{ECas3} we already made the first step in the asymptotic expansion by substituting the summations by integrations. The next step is to expand in $\M$ the $A_{m,m'}$, which are given by Eqs. \Ref{ATE} and \Ref{ATM}.  We start from the factors $R^{\delta_{\rm TE}}$ and $R^{\delta_{\rm TM}}$. Here we have simply to insert the uniform asymptotic expansions of the modified Bessel functions. We note that the exponential factors are the same as for hard boundaries since they cancel in the $\Om$-dependent denominators. We obtain
\be\label{RdTE}R^{\delta_{\rm TE}}\raisebox{-4pt}{${\sim\atop \ep\to0}$}
\frac{e^{\eta_0}}{\pi} \  r^{(\delta_{\rm TE})}\left(1+P^{(\delta_{\rm TE})} \sqrt{\ep}+
Q^{(\delta_{\rm TE})} \ep +\dots\right)
\ee
and
\be\label{RdTM}R^{\delta_{\rm TM}}\raisebox{-4pt}{${\sim\atop \ep\to0}$}
\frac{e^{\eta_0}}{\pi} \  r^{(\delta_{\rm TM})}\left(1+P^{(\delta_{\rm TM})} \sqrt{\ep}+
Q^{(\delta_{\rm TM})} \ep +\dots\right)
\ee
with
\bea\label{rr}r^{(\delta_{\rm TE})}&=&\frac{1}{1+t/\Om_L} ,\nn \\
r^{(\delta_{\rm TM})}&=&\frac{-1}{1+ty^2/\Om_L},
\eea
where we introduced the notation
\be\label{y}y=\sqrt{1-\tau^2}\sin\al.
\ee
Details on the calculation and explicit expressions for $P^{(\delta_{\rm T*})}$ and $Q^{(\delta_{\rm T*})}$ are given in the Appendix A, Eqs.\Ref{PQ}. The factor $\eta_0$ in the exponential will be considered below together with the corresponding one from $\A_{m+m'}$. 

We continue with the expansion of $\A_{m+m'}$. In the case of hard boundary conditions this was due to \Ref{Kh} simply the asymptotic expansion of the Bessel function. Here we have to consider the integral representation \Ref{AT}. However, since it is very close to a known integral representation of the Bessel function we can use standard methods, namely a saddle point expansion. This done in Appendix B where also the result is stated. It consists of a factor $\phi$, Eqs.\Ref{phi1}-\Ref{phi3}, which is different for the models considered and factors which are the same as for hard boundary conditions in \meins. These factors are collected in $\psi$, Eq.\Ref{Apsi}. With all these factors at hand we obtain for the functions $ A_{m+n,m+n'}$ entering $\M$, \Ref{M}, 
\be\label{A3}A_{m+n,m+n'}=
\sqrt{\frac{\ep}{4\pi t}} \ \ e^{-{\eta^{\rm as}}} \
r^{(\rm cyl)}r^{\rm (plane)} \
 \left(1+\sqrt{\ep} \ a_{n,n'}^{(1/2)} +\ep \ a_{n,n'}^{(1)}  +\dots\right) \ ,
\ee
where the factor in the exponential is the same as in \meins, Eq.(57),
\be\label{etaas}
\eta^{\rm as}=2t+(n-n')^2.
\ee
Formula \Ref{A3} has the same structure as Eq.(56) in \meins ~with the only differences that the functions $a_{n,n'}^{(1/2)}$ and $ a_{n,n'}^{(1)}$ are different (they are displayed in Appendix C)  and that the additional factors $r^{(\rm cyl)}$ and $r^{\rm (plane)}$ are present. These are in the given variables just the reflection coefficients for the corresponding boundary conditions on planes whereby $r^{(\rm cyl)}$ originates from $R^{\delta_{\rm TE}}$ in \Ref{ATE} and $R^{\delta_{\rm TM}}$ in \Ref{ATM}, i.e., from
the cylinder, and $ r^{\rm (plane)}$ originates from $\tilde{d}_{\om,k_3}$ in $A_{m+m'}$, \Ref{AT}, i.e., from the plane. 
There are four  expressions,
\be\label{rdp}r^{\rm (cyl,plane)}=\left\{\begin{array}{cl}
\frac{\Om_L}{\Om_L+t},\qquad & \mbox{for plasma sheet, TE mode},\\ [9pt]
-\frac{\Om_L}{t   y^2+\Om_L} \qquad & \mbox{for plasma sheet, TM mode},\\ [9pt]
\frac{\sqrt{\om_L^2+t^2}-t}{t+\sqrt{\om_L^2+t^2}} \qquad & \mbox{for dielectric, TE mode} , \\ [9pt]
\frac{t\left(\sqrt{\om_L^2+t^2}-t\right) y^2-\om_L^2}{\om_L^2+t \left(t+\sqrt{\om_L^2+t^2}\right)   y^2}& \mbox{for dielectric, TM mode} ,
\end{array}\right.
\ee
whereby $r^{(\rm cyl)}$ can be equal to the first two and $r^{(\rm plane)}$ to any of them. 

In \Ref{rdp}  we used the notation $y$, Eq.\Ref{y}.
The variables $\tau$ and $\al$ enter the reflection coefficients just in this combination. In the leading order for small $\ep$ this is the only dependence on $\tau$ and $\al$ and one of these integrations can be carried out even in the TM case. In 
higher orders in $\ep$, from the expressions $a_{n,n'}^{(1/2)}$ and $a_{n,n'}^{(1)}$, a dependence  on both variables comes in. However, the dependence on $\tau$ is polynomial and can be cared for by means of the simple formula  
\bea\label{Hilfsf}
\int_0^{\pi/2}\frac{d\al}{\pi/2}\int_0^1d\tau \ \tau^{2n} f(\sqrt{1-\tau^2}\sin\al)
= \
\frac{\Gamma(n+\frac12)}{\sqrt{\pi}\Gamma(n+1)} \int_0^1 dy \ (1-y^2)^{n} f(y) \ .
\eea
Next we insert \Ref{A3} into $\cal M$, \Ref{M},
and after a reexpansion we obtain 
\beao
{\cal M}&=&\left(\frac{\ep}{4\pi t}\right)^\frac{s+1}{2} \ \ e^{-2(s+1)t-\eta_1}
\ r^{(\rm cyl)}r^{\rm (plane)} \ \M^{\rm as}
\nn
\eeao
with
\bea        \label{Mas}
\M^{\rm as}&=&\left(1+\sqrt{\ep}\ \sum_{i=0}^s  a_{n_i,{n}_{i+1}}^{(1/2)}
\right.  \\ && \left. +\ep\left(\sum_{0\le i <  j\le s}  a_{n_i,{n}_{i+1}}^{(1/2)}a_{n_j,{n}_{j+1}}^{(1/2)}
 +\sum_{i=0}^s  a_{n_i,{n}_{i+1}}^{(1)}
 \right)+\dots\right)   \nn
\eea
and
\be\label{eta1}\eta_1= \sum_{i=0}^{s}\left(n_i-n_{i+1}\right)^2,
\ee
where, in order to include all contributions into the sum signs, we have  to put $n_0=n_{s+1}=0$ formally.
With these formulas the Casimir energy $E_{\rm Cas}$, \Ref{ECas3}, becomes
\bea\label{ECas4}%
E_{\rm Cas}&=& 
\frac{-\ep^{-5/2}}{2\pi}\sum_{s=0}^\infty\frac{1}{s+1}
\int_0^\infty\frac{dt}{t}\frac{t^{5/2}e^{-2t(s+1)}}{\sqrt{4\pi}}
\int_0^{\pi/2}\frac{d\al}{\pi/2}\int_0^1d\tau \ 
\nn \\ && ~~~~~~
    \times
\int_{-\infty}^\infty \frac{dn_1}{\sqrt{\pi}} \ \dots \int_{-\infty}^\infty \frac{dn_s}{\sqrt{\pi}} \ \  e^{-\eta_1} \ \ 
\ \left( r^{(\rm cyl)}r^{\rm (plane)} \right)^{s+1}\ \M^{\rm as} \ .
\eea
This expression for the Casimir energy is to some extend the final formula for the short separation expansion because the remaining integrations cannot be carried out in an explicit form (except for that over the $n_i$). It represents the generalized Lifshitz formula beyond PFA. In the following we discuss its basic features. 

First of all, the dimensions are restored by multiplying the whole expression be $R^{-2}$. With $\ep^{-5/2}R^{-2}=\frac{1}{L^2}\sqrt{\frac{R}{L}}$ the dimensional factor known from the PFA is restored. The dimensions inside the above expressions are restored by means of $\Om_L\to \Om L$ and $\om_L\to\om_p L$.
Further, it is obvious that the limes of hard boundary conditions is recovered for $\Om\to\infty$ and $\om_p\to\infty$ since we arrive just to the same expressions as in \meins. 

The first nontrivial result which follows from \Ref{ECas4} is a confirmation of the PFA for semitransparent boundaries derived in \cite{BORDAG2006D}. It comes about from the leading order in $\ep$, i.e., with $\M^{\rm as}\to 1$ in \Ref{ECas4}.
With this, there is no dependence on the $n_i$ besides in $\eta_1$ and the integrations over the $n_i$ can be carried out using Eq.(66) in \meins ~delivering a factor of $(s+1)^{-1/2}$. Next we observe that there is no dependence on $\tau$ so that formula \Ref{Hilfsf} can be used with $n=0$. Finally, the summation over $s$ is expressed in terms of a polylogarithm, $\Li_s(z)=\sum_{n=1}^\infty z^n/n^s$, and we arrive at
\bea\label{EPFA}
E_{\rm Cas}^{\rm PFA}&=&\frac{-1}{4\pi^{3/2} L^2}\sqrt{\frac{R}{L}}
\int_0^\infty dt \ t^{3/2} \int_0^1 dy \ \Li_{3/2}\left[ r^{(\rm cyl)}r^{\rm (plane)} \ e^{-2t} \right]
\eea
for one of the scalar problems with $r^{(\rm cyl)}$ and $r^{\rm (plane)}$ given by Eqs.\Ref{rdp}. Up to  differences in notations this coincides with  Eq.(33) in \cite{BORDAG2006D}. 

In order to represent the result in a more instructive manner we rewrite \Ref{ECas4} in the form
\bea\label{Efg}
E_{\rm Cas}&=&
-\frac{\pi^3}{1920\sqrt{2}L^2}\sqrt{\frac{R}{L}}\left(f_0(\Om L,\om_p L)
+\frac{L}{R} \ f_1^{\rm h} \ 
f_1(\Om L,\om_p L)+\dots\right),
\eea
where the factor in front is the Casimir energy for hard boundary conditions (Dirichlet for TE  and and Neumann for TM, both give the same) in PFA. The
function $f_0$  represents the relative decrease of the Casimir energy in the considered model in PFA. The function $f_1$ is the first contribution beyond PFA. It is also written relative to the hard boundary case for which we have \meins
\be             
\label{f1h}
f_1^{\rm h}=\left\{\begin{array}{ll}
\frac{7}{36} \quad & \mbox{for TE},
\\ [9pt]
\frac{7}{36}-\frac{40}{3\pi^2} \quad & \mbox{for TM}.
\end{array}
\right.
\ee
Let us consider the PFA, i.e., the function $f_0$, in more detail.
It  follows from Eq.\Ref{EPFA} with $r^{(\rm cyl)}$ and $r^{\rm (plane)}$  for the models formulated in section 2,
\bea\label{f0g0}
f_0^{(\delta\delta)_{\rm TE}}(\Om L)&=&\frac{480\sqrt{2}}{\pi^{9/2}}
\int_0^\infty dt \ t^{3/2} \ 
\Li_{3/2}\left[\frac{\exp(-2t)}{(1+t/\Om L)^2}\right] \ ,
\nn \\ 
f_0^{(\delta\delta)_{\rm TM}}(\Om L)&=&\frac{480\sqrt{2}}{\pi^{9/2}}
\int_0^\infty dt \ t^{3/2} \int_0^1 dy \ 
\Li_{3/2}\left[\frac{\exp(-2t)}{(1+ty^2/\Om L)^2}\right] \ ,
\nn \\
f_0^{(\ep\delta)_{\rm TE}}(\Om L,\om_pL)&=&\frac{480\sqrt{2}}{\pi^{9/2}}
\int_0^\infty dt \ t^{3/2}  \ 
\Li_{3/2}\left[\frac{\sqrt{(\om_pL)^2+t^2}-t}{t+\sqrt{(\om_pL)^2+t^2}} \ \frac{\exp(-2t)}{1+t/\Om L}\right],
\nn \\
f_0^{(\ep\delta)_{\rm TM}}(\Om L,\om_pL)&=&\frac{480\sqrt{2}}{\pi^{9/2}}
\int_0^\infty dt \ t^{3/2} \int_0^1 dy \ 
\\ &&
\Li_{3/2}\left[\frac{t\left(t-\sqrt{(\om_pL)^2+t^2}\right) y^2+(\om_pL)^2}{(\om_pL)^2+t \left(t+\sqrt{(\om_pL)^2+t^2}\right)   y^2} \ \frac{\exp(-2t)}{1+ty^2/\Om L}\right].
\nn
\eea
The limiting value for hard boundary conditions which appears for  $\Om L\to\infty$, $ \om_pL\to\infty$ is $f_0\to 1$ for all four functions. In that case $L<<\frac{1}{\Om}$, $L<<\frac{1}{\om_p}$ and $L<<R$ holds.

The opposite limiting case is $\frac{1}{\Om}<<L<<R$ and $\frac{1}{\om_p}<<L<<R$. It corresponds to small separation as compared with the plasma wave length. This is the nonretarded regime or the case when the Casimir forces turn into the van der Waals forces. 
It was already mentioned in \cite{BORDAG2006D}. 

\begin{figure}\unitlength1cm 
 \begin{picture}(14,5)
 \put(0,0){ \includegraphics[width=7cm]{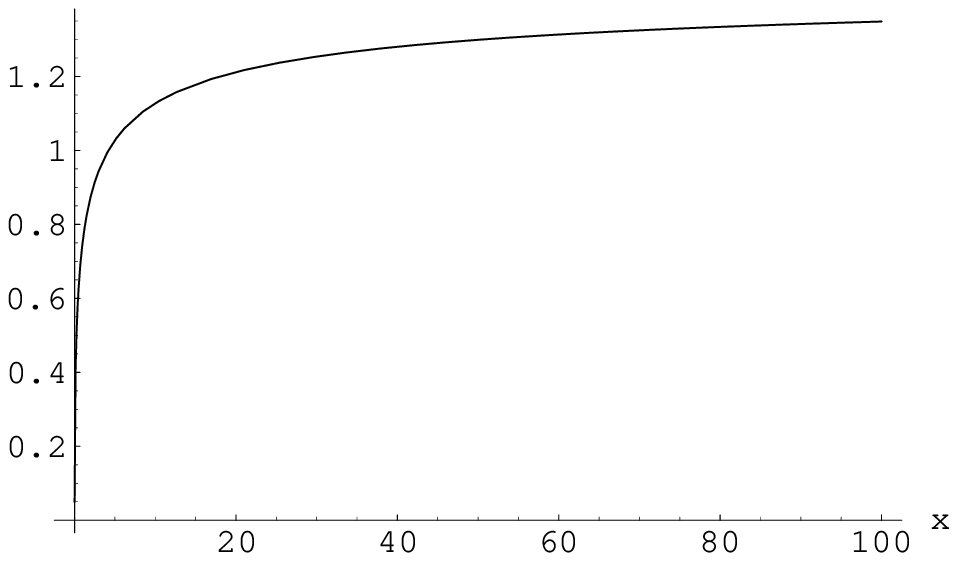}}
 \end{picture}
 \caption{The function $\tilde{f}^{(\ep\delta)}\left(x\right)$ appearing in \Ref{smallTM}
} \label{ftilde}
\end{figure}

To obtain that limiting case, for the TE cases one simply expands the polylogarithm formally for small $L$ and the remaining integration then gives
\bea\label{smallTE}
f_0^{(\delta\delta)_{\rm TE}}(\Om L)&\raisebox{-4pt}{${\sim\atop \Om L\to 0}$}&\frac{840}{\pi^4} \ (\Om L)^2 \ ,
\nn \\ [9pt]
f_0^{(\ep\delta)_{\rm TE}}(\Om L,\om_p L)&\raisebox{-4pt}{${\sim\atop \Om L\sim \om_p L\to 0}$}&\frac{420}{\pi^4} \ \om_p\Om L^2 \ .
\eea
In this limit the TE mode is subleading. The leading contribution comes from the TM mode. Here it is impossible to simply expand the arguments of the polylogarithm because then the remaining integrations become singular. Instead on has to make the substitution either $y\to y\om_p L$ or  $y\to y\sqrt{\Om L}$ after which the limit $L\to 0$ can be taken,
\bea\label{smallTM}
f_0^{(\delta\delta)_{\rm TM}}(\Om L)&\raisebox{-4pt}{${\sim\atop \Om L\to 0}$}&
\frac{840\sqrt{2}}{\pi^{9/2}} \ \sqrt{\Om L} \ \tilde{f}^{(\delta\delta)} ,
\nn \\ [9pt]
f_0^{(\ep\delta)_{\rm TM}}(\Om L,\om_p L)&\raisebox{-4pt}{${\sim\atop \Om L\sim \om_p L\to 0}$}&
\frac{840\sqrt{2}}{\pi^{9/2}} \ \sqrt{\Om L} \ \tilde{f}^{(\ep\delta)}\left(\frac{\om_p^2 L}{\Om}\right) \ ,
\eea
with
\bea \tilde{f}^{(\delta\delta)}&=& \int_0^\infty dt \ t^{3/2} \int_0^\infty dy \ 
\Li_{3/2}\left[\frac{\exp(-2t)}{(1+ty^2)^2}\right]
\nn \\ &=& 0.254,
\eea
and
\bea\label{fti}
\tilde{f}^{(\ep\delta)}\left(\frac{\om_p^2 L}{\Om}\right)&=&
 \int_0^\infty dt \ t^{3/2} \int_0^\infty dy \ 
\Li_{3/2}\left[\frac{\exp(-2t)}{(1+ty^2\om_p^2L/\Om)(1+2t^2y^2)}\right].
\eea
The function $\tilde{f}^{(\ep\delta)}\left(\frac{\om_p^2 L}{\Om}\right)$ is shown in Fig.\ref{ftilde}. Its limiting values are 
$\tilde{f}^{(\ep\delta)}\left(x\right) \raisebox{-4pt}{${\sim\atop x\to 0}$} 1.62\sqrt{x}$ and 
$\tilde{f}^{(\ep\delta)}\left(x\right) \raisebox{-4pt}{${\sim\atop x\to \infty}$} 1.39$.
\begin{figure}[h]\unitlength1cm
 \begin{picture}(14,5)
 \put(0,0){ \includegraphics[width=7cm]{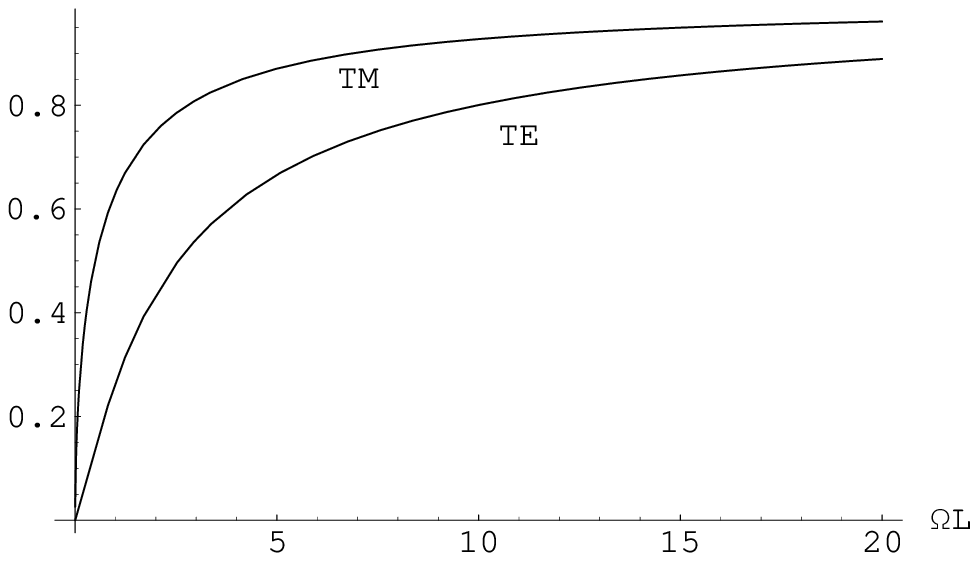}}
 \end{picture}
 \caption{The function $f_0^{\delta\delta}(\Om L)$ representing the decrease of the Casimir energy at small separation relative to the hard boundary case for the TE and for the TM modes in the interaction of the plasma shell cylinder with a flat plasma shell in PFA.} \label{fig1}
\end{figure}

\begin{figure}\unitlength1cm
 \begin{picture}(14,5)
 \put(0,0){ \includegraphics[width=7cm]{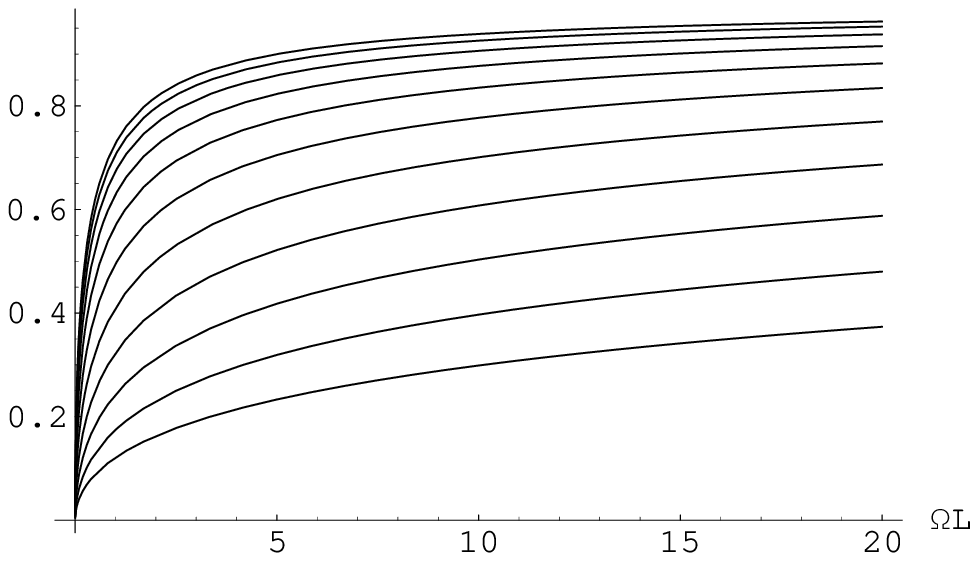}}
 \end{picture}
 \caption{The function $f_0^{(\ep\delta)_{\rm TM}}(\Om L,\om_pL)$ with $\om_p=\sqrt{\Om\la/L}$ for $\la=10^j$ from $j=-2$ (lower curve) till $j=2$ (upper curve) in equal steps.}
\label{fig3}
\end{figure}

The functions $f_0$ for the $(\delta\delta)$-case, i.e., for the interaction of the cylindrical plasma sheet with a flat plasma sheet, are shown in Fig.\ref{ftilde}. For large argument they go to unity which is the hard boundary limit. For smaller argument when the sheets become more transparent they decrease whereby the function for the TE case decreases faster. The behavior at the origin is given by the upper lines in Eqs.\Ref{smallTE} and \Ref{smallTM}. The corresponding functions for the $(\ep\delta)$-model depend on two parameters. Their general behavior is similar to the $(\delta\delta)$-case and as an example we show in Fig.\ref{fig3} the function $f_0^{(\ep\delta)_{\rm TM}}(\Om L,\om_pL)$ for several values of the ratio of its arguments. Also for these function the limiting value for large arguments, i.e., for hard boundaries is unity and the limit for small arguments is given by the lower lines in Eqs.\Ref{smallTE} and \Ref{smallTM}.

The first contribution beyond PFA is given by the functions $f_1$ in Eq.\Ref{Efg}. These appear  from the contributions proportional to $\sqrt{e}$ and to $\ep$
in the representation \Ref{ECas4} of the energy with $a^{(1/2)}_{n,n'}$ and  $a^{(1)}_{n,n'}$, Eqs.\Ref{a12ddTE} till \Ref{a1epdTM}, inserted into $\M^{\rm as}$, \Ref{Mas}. The next step is then to carry out the integrations over the $n_i$. These integrations are Gaussian since the exponential $\eta_1$ is quadratic in the $n_i$ and $\M^{\rm as}$ is polynomial. As seen from the explicit formulas, $a^{(1/2)}_{n,n'}$ is odd in the $n_i$ hence the contribution  proportional to $\sqrt{\ep}$ vanishes. So we are left with the integrations of the  terms proportional to $\ep$. These integrations can be carried out in the same manner as in \meins. However, the results are too big as to be displayed here. After that  formula \Ref{Hilfsf} can be used since the appearing expressions are polynomial in $\tau$. Again, the  integration over $y$ remaining in \Ref{Hilfsf}  can be carried out trivially in the TE case but not in the TM case. The next step is  to rewrite the summation over $s$ in terms of polylogarithms. After that the expressions for the functions $f_1$ become simpler. They are represented by formulas similar to that for the $f_0$, Eqs.\Ref{f0g0}, however more lengthy and for the $(\delta\delta)$-model the explicit expressions are displayed in Appendix D, Eq.\Ref{f0g0D}.

\begin{figure}\unitlength1cm
 \begin{picture}(14,5)
 \put(0,0){ \includegraphics[width=7cm]{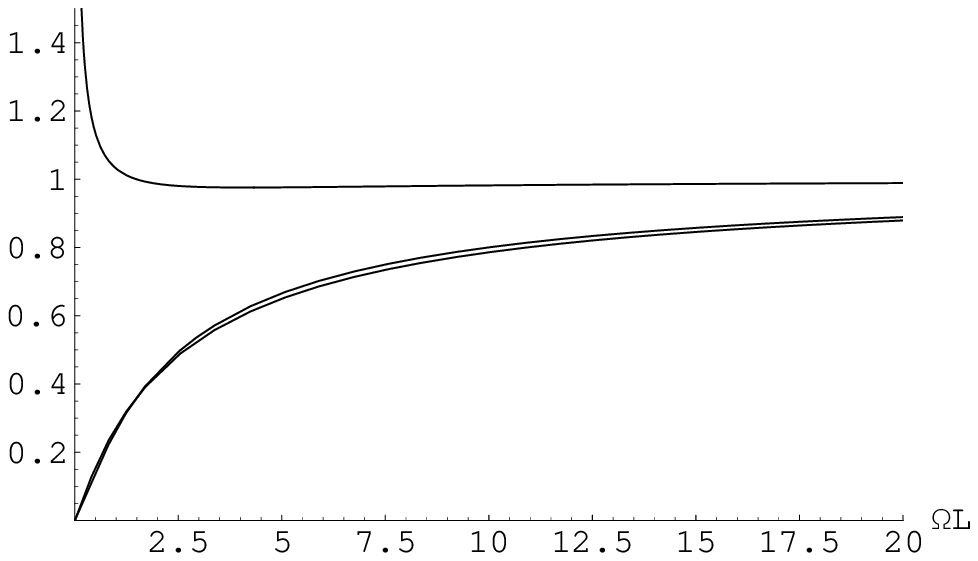}}
 \end{picture}
 \caption{The function $f_0^{(\delta\delta)_{\rm TE}}(\Om L)$ and $f_1^{(\delta\delta)_{\rm TE}}(\Om L)$  (lower two curves) and their ratio $f_1^{(\delta\delta)_{\rm TE}}(\Om L)/f_0^{(\delta\delta)_{\rm TE}}(\Om L)$ (upper curve) for the TE-mode in the $(\delta\delta)$-model.}
\label{figddE}
\end{figure}

It is interesting to note that the functions $f_0$ and $f_1$ are numerically quite close to each other. For the TE-mode in the $(\delta\delta)$-model both functions are shown in Fig.\Ref{figddE}. Only for small values of the argument $\Om L$ there is a significant difference. It comes about because in this case it is impossible to expand the integrand for the function for $f_1$ simply in powers of $\Om L$. Instead one has to make a rescaling of the integration variable and the behavior comes out to be
\be\label{sm}
f_1^{(\delta\delta)_{\rm TE}}(\Om L)
\raisebox{-4pt}{${\sim\atop \Om L\to0}$}
\frac{720\sqrt{2}}{7\pi^{7/2}} \ \left(\Om L\right)^{3/2}.
\ee
This behavior is different from the corresponding one in PFA (upper line in Eq.\Ref{smallTE}). 

\begin{figure}\unitlength1cm
 \begin{picture}(14,5)
 \put(0,0){ \includegraphics[width=7cm]{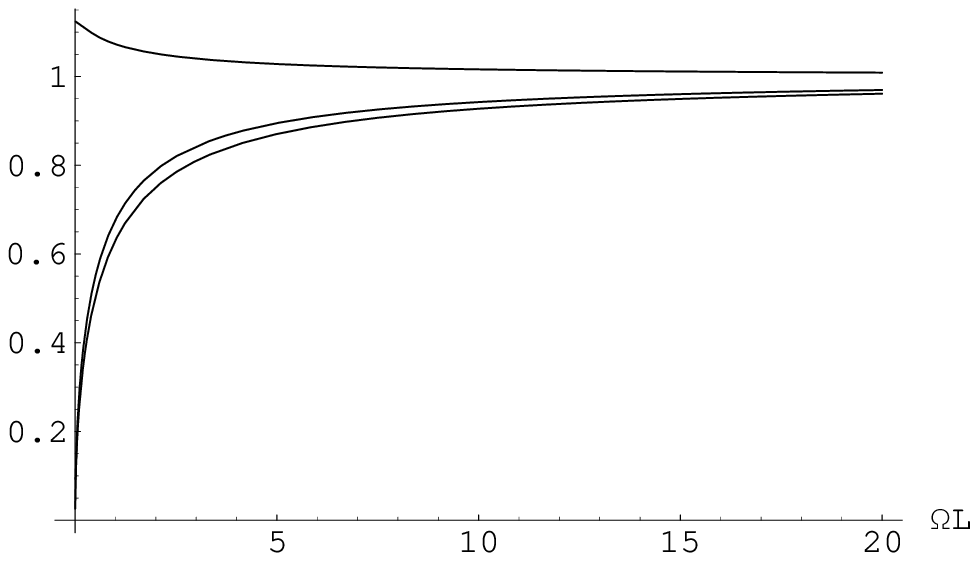}}
 \end{picture}
 \caption{The function $f_0^{(\delta\delta)_{\rm TE}}(\Om L)$ and $f_1^{(\delta\delta)_{\rm TE}}(\Om L)$  (lower two curves) and their ratio $f_1^{(\delta\delta)_{\rm TE}}(\Om L)/f_0^{(\delta\delta)_{\rm TE}}(\Om L)$ (upper curve) for the TM-mode in the $(\delta\delta)$-model. The function $f_1^{(\delta\delta)_{\rm TE}}(\Om L)$ takes always the larger values as compared to $f_0^{(\delta\delta)_{\rm TE}}(\Om L)$.}
\label{figddM}
\end{figure}

For the TM-mode in the $(\delta\delta)$-model both functions are shown in Fig.\Ref{figddM}.
Here the behavior for small argument is the same as in PFA and can be calculated by the same substitution. The result is
\be
f_1^{(\delta\delta)_{\rm TM}}(\Om L)  \raisebox{-4pt}{${\sim\atop \Om L\to0}$}
0.92 \sqrt{\Om L} .
\ee
As a consequence, and as seen from the upper curve in Fig.\Ref{figddM}, the ratio 
$\frac{f_1^{(\delta\delta)_{\rm TE}}(\Om L)}{f_0^{(\delta\delta)_{\rm TE}}(\Om L)}$ of the two curves takes finite values for all $\Om L$.  We mention that like in PFA the TM mode dominates at small separation.

\begin{figure}\unitlength1cm
 \begin{picture}(14,5)
 \put(0,0){ \includegraphics[width=7cm]{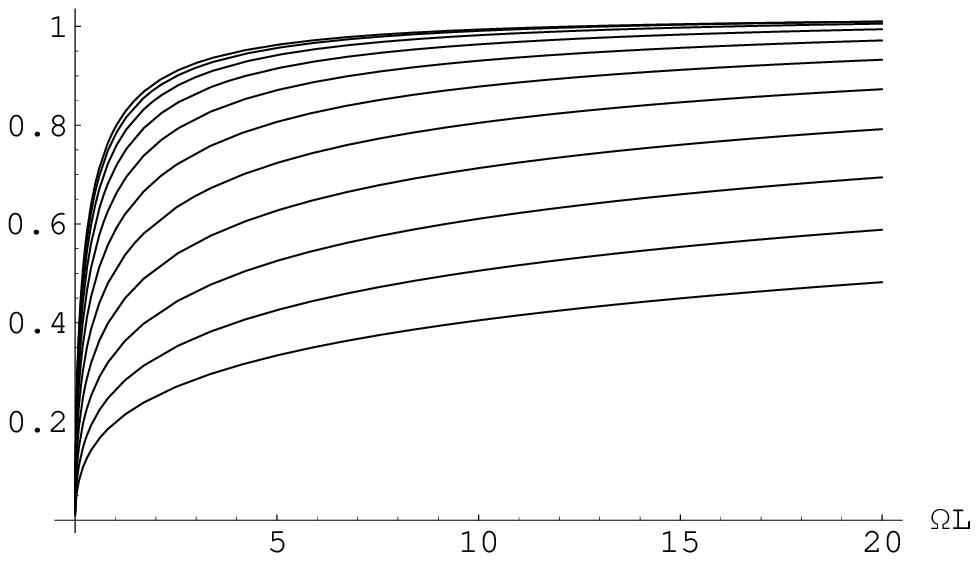}}
 \end{picture}
 \caption{The function $f_1^{(\ep\delta)_{\rm TM}}(\Om L,\om_pL)$ with $\om_p=\sqrt{\Om\la/L}$ for $\la$ from $\la=0.01$ (lower curve) till $\la=100$ (upper curve).}
\label{figedTM}
\end{figure}

\begin{figure}\unitlength1cm
 \begin{picture}(14,5)
 \put(0,0){ \includegraphics[width=7cm]{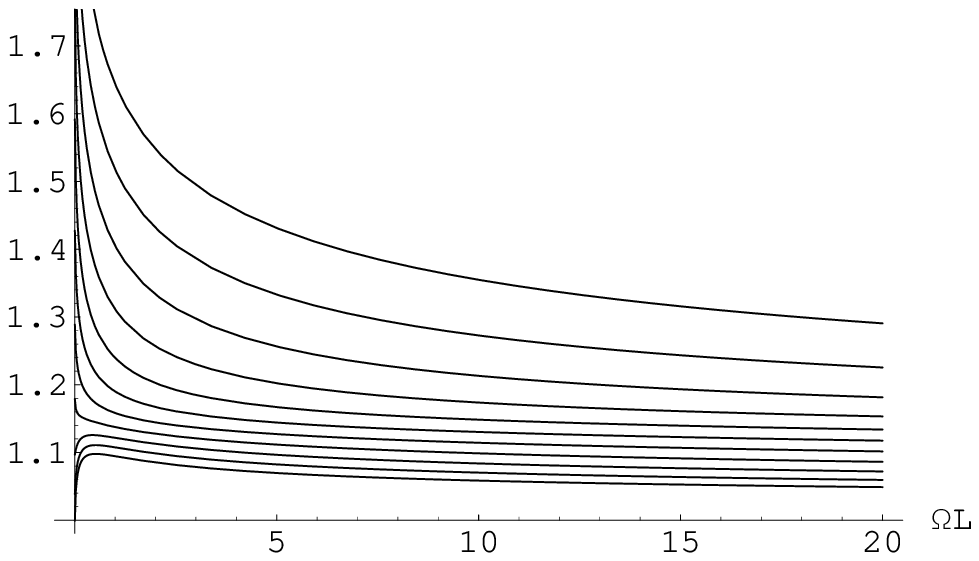}}
 \end{picture}
 \caption{The  ratio  $f_1^{(\ep\delta)_{\rm TM}}(\Om L,\om_pL)/f_0^{(\ep\delta)_{\rm TM}}(\Om L,\om_pL)$ in the $(\ep\delta)$-model  with $\om_p=\sqrt{\Om\la/L}$ for $\la$ from $\la=0.01$ (lower curve) till $\la=100$ (upper curve).}
\label{figedTMratio}
\end{figure}

Finally we discuss the results for the $(\ep\delta)$-model. Again, the functions $f_1$ are very close to the functions $f_0$ in PFA. So we restrict us to show in Fig.\ref{figedTM} the function $f_1^{(\ep\delta)_{\rm TM}}(\Om L,\om_pL)$ for several values of the ratio of their arguments.
It is more instructive to consider the ratio of the functions, $f_1^{(\ep\delta)_{\rm TM}}(\Om L,\om_pL)/f_0^{(\ep\delta)_{\rm TM}}(\Om L,\om_pL)$, in Fig.\ref{figedTMratio}. Again, the function $f_1$ is always a bit larger than the function $f_0$ keeping a finite ratio.
%
%
\section{Conclusions}
In the forgoing sections we have derived the generalization of the Lifshitz formula for the interaction of a cylindrical plasma sheet with a flat plasma sheet ($(\delta\delta)$-model) and with a dielectric half space ($(\ep\delta)$-model) in PFA and in the first approximation beyond. The limiting values for hard boundary conditions and for a separation small as compared with the plasma wave lengths of the model (which corresponds to the nonretarded regime) are considered in detail. The previously known results were reproduced. The results in PFA and beyond are represented in form of functions $f_0$ and $f_1$ describing the deviation from the hard boundary limes, numerically computed and represented in a number of figures. In general words the outcome is that these functions are very close to each other, i.e., the functions describing the ratios of the two functions $f_1$ and $f_0$ are close to unity. This allows a simpler representation of the result for the needs of practical calculations, namely as a pure superposition of PFA and the first correction beyond (as calculated so far in \meins ~and \cite{GIES2006}) and the reduction function taking care of the frequency dispersion (or transparency) of the boundaries which can be calculated for flat boundaries. In terms of formulas that would imply in a representation like \Ref{Efg} simply to take the function $f_0$ in place of the function $f_1$. It can be expected that this property holds also if one considers for example a dielectric medium with a dispersion given by optical data. 

Open questions are still an extension of the method to a sphere in front of a plane and for accounting for the non separation of the modes of the electromagnetic field. However, for small separations the TM mode will prevail anyway and for large separations which is equivalent to the hard boundary limes both modes are additive (for the wave guide geometry). So it can be expected that for practical calculations the above results will be sufficient at present time.

\section*{Acknowledgements}
This work was supported by the research funding from the EC's Sixth Framework Programme within the STRP project "PARNASS" (NMP4-CT-2005-01707).
\setcounter{equation}{0}\renewcommand{\theequation}{A.\arabic{equation}}
\section*{Appendix A}
In this appendix we calculate the uniform asymptotic expansions of the factors $R^{\delta_{\rm TE}}$ and $R^{\delta_{\rm TM}}$ appearing in Eqs.\Ref{RdTE} and \Ref{RdTM}. We insert the well know uniform asymptotic expansion of the modified Bessel functions into these formulas. The exponential factors in the $\Om$-dependent denominators cancel and that from the quotient of Bessel functions will be considered later combined with that from the $\A_{m+m'}$. For the moment we denote then by $\eta_0$. We introduce the intermediate notations $z=\rho/m$ and $z'=\rho/m'$ and obtain
\be\label{67}
R^{\delta_{\rm TE}}=\frac{e^{\eta_0}}{\pi} \ \frac{1}{1+\frac{\sqrt{m^2+\rho^2}}{\Om}} \ 
\sqrt{\frac{m}{m'}} \ \left(\frac{1+z'^2}{1+z^2}\right)^\frac14 \ \left(1+\frac{u_1(z)}{m}+\frac{u_1(z')}{m'}+\dots\right),
\ee
where $u_1(z)=(3(1+z^2)-5)/(24(1+z^2)^{3/2})$ results from the first Debye polynomial. 
In the same way we obtain for the TM case
\be\label{68}
R^{\delta_{\rm TM}}=\frac{e^{\eta_0}}{\pi} \ \frac{1}{1+\frac{\om^2}{\Om \sqrt{m^2+\rho^2}}} \ 
\sqrt{\frac{m}{m'}} \ \left(\frac{1+z^2}{1+z'^2}\right)^\frac14 \ \sqrt{\frac{z'}{z}} \  \left(1+\frac{v_1(z)}{m}+\frac{v_1(z')}{m'}+\dots\right)
\ee
with $v_1(z)=(-9(1+z^2)+7)/(24(1+z^2)^{3/2})$.
Now by make the substitution \Ref{varsub} and expansion with respect to $\ep$ delivers the expressions stated in Eq.\Ref{RdTE} and \Ref{RdTM} with
\bea\label{PQ}
P^{\delta_{\rm TE}}&=&-\frac{(n (t-\Om_L)+n' (\Om_L+t)) \tau }{\sqrt{t} (\Om_L+t)}
,\nn \\
Q^{\delta_{\rm TE}}&=&\frac{1}{{12 t (\Om_L+t)^2}}
\Big(\left(-6 \left(3 \tau ^2-2\right) n^2-12 n' \tau ^2 n-5 \tau ^2
\right. \nn \\ && \left.
+6 n'^2 \left(5 \tau ^2-2\right)+3\right) \Om_L^2+2 t \left(6 \left(5 \tau ^2-2\right) n'^2
\right. \nn \\ && \left.
-\left(18 n^2+5\right) \tau ^2+3\right) \Om_L+t^2 \left(6 \left(5 \tau ^2-2\right) n^2+12 n' \tau ^2 n
\right. \nn \\ && \left.
-5 \tau ^2+6 n'^2 \left(5 \tau ^2-2\right)+3\right)\Big)
,\nn \\
P^{\delta_{\rm TM}}&=&
\frac{\tau  \left(n t y^2+n' t y^2-n \Om_L+n' \Om_L\right)}{\sqrt{t} \left(t y^2+\Om_L\right)}
,\nn \\
Q^{\delta_{\rm TM}}&=&
\frac{1}{12 t \left(t y^2+\Om_L\right)^2}
\Big(t^2 \left(-6 \left(3 \tau ^2-2\right) n^2+12 n' \tau ^2 n+7 \tau ^2
\right. \nn \\ && 
\left.-6 n'^2 \left(3 \tau ^2-2\right)-9\right) y^4-2 \Om_L t \left(6 \left(3 \tau ^2-2\right) n'^2
\right. \nn \\ && \left.
+\left(18 n^2-7\right) \tau ^2+9\right) y^2+\Om_L^2 \left(6 \left(5 \tau ^2-2\right) n^2-12 n' \tau ^2 n
\right. \nn \\ && \left.
+7 \tau ^2-6 n'^2 \left(3 \tau ^2-2\right)-9\right)\Big)
. \eea
We took into account that $\Om\ep\equiv\Om_L$ and $\om_p \ep\equiv\om_L$ must not be considered as  small. 

\setcounter{equation}{0}\renewcommand{\theequation}{B.\arabic{equation}}
\section*{Appendix B}
In this appendix we calculate the asymptotic expansion of the expressions $\A_{m+m'}$ in \Ref{ATE} and \Ref{ATM}. For this we use a saddle point expansion. Let
\be\label{sp}
\A=\int_0^\infty d\theta \  g(\theta) \ e^{-\la h(\theta)}
\ee
be the integral to be considered and $\la$ a big parameter. Assuming the function $h(\theta)$ has a minimum for some $\theta_0\in(0,\infty)$, the expansion for $\la\to\infty$ is
\be\label{sp1}\A\sim\sqrt{\frac{\pi }{2\la h_2}} \ e^{-\la \eta(z)} \ \phi
\ee
with
\be
\label{phi}  \phi=
g_0
+\left(\left(\frac{5h_3^2}{24h_2^3}
-\frac{h_4}{8h_2^2}\right)g_0-\frac{h_3g_1}{2h_2^2}+\frac{g_2}{2h_2}\right)\frac{1}{\la}
+\dots,
\ee
where $g_i=g^{(i)}(\theta_0)$ and $h_i=h^{(i)}(\theta_0)$ are the derivatives in the minimum, $h'(\theta_0)=0$ and $\eta(z)$ is known from the uniform asymptotic expansion of the modified Bessel functions.

For applying this expansion to $\A_{m+m'}$, we have 
first to rewrite  Eq.\Ref{AT} in the form
\be\label{AT1}\A_{m+m'}=\frac12 \int_0^\infty d\theta \ 
\ 2\rho\cosh\theta \ \tilde{d}_{\om,\ga} \ e^{-2a\rho\cosh\theta-(m+m')\theta}
\ee
and then to identify
\bea\label{}g(\theta)&=& 2\rho\cosh\theta \ \tilde{d}_{\om,\ga} \ , \nn \\
h(\theta)&=&\ep\left(2a\rho\cosh\theta+(m+m')\theta\right).
\eea
With the intermediate notations 
\be\label{}\la=m+m' \ , \quad z=\frac{2\rho(1+\ep)}{m+m'}
\ee
the expansion reads
\bea\label{sp2}\A_{m+m'}&\sim&
\sqrt{\frac{\pi}{2\la}} \ \frac{e^{-\la \eta(z)}}{(1+z^2)^{1/4}}
 \Bigg\{ g_0
  \\ &&
 +\frac{1}{\la} \ \left[\frac{5-3(1+z^2)}{24(1+z^2)^{3/2}} g_0
+\frac{g_1}{2(1+z^2)}+\frac{g_2}{2(1+z^2)^{1/2}}
\right]+\dots
\Bigg\}\nn \ .
\eea
We remark that the function $h(\theta)$ has just the same structure as in the known integral representation of the modified Bessel function and that the whole difference is in the function $g(\theta)$.

Next we insert the substitution \Ref{varsub} and reexpand taking into account that $\A_{m+m'}$ enters $\M$ in \Ref{M} in fact with  $m\to m+n$ and $m'\to m+n'$.  Using \Ref{dtallg} for $\tilde{d}_{\om,\ga}$ in $g(\theta)$
we obtain
\be\label{}\A_{m+m'}\sim\sqrt{\frac{\pi\ep}{4t}} \ \frac{1}{\tau}  \ \phi \ ,
\ee
where for $\phi$ in dependence on the case considered one has to insert one of the following,
\bea\label{phi1}
\phi^{\delta_{\rm TE}}&=&
\frac{\Om_L}{\Om_L+t}-\frac{(n+n') \Om_L \sqrt{t} \tau  }{(\Om_L+t)^2}\sqrt{\ep}
 \\ &&
+\frac{1}{48 t (\Om_L+t)^3}\Big(\Om_L \left(\left(5 \tau ^2-3\right) \Om_L^2+2 t \left(12 \left(\tau ^2-1\right) n^2+24 n' \left(\tau ^2-1\right) n
\right. \right.\nn \\ && \left.\left.
+24 t \tau ^2+11 \tau ^2+12 n'^2 \left(\tau ^2-1\right)-9\right) \Om_L+t^2 \left(24 \left(3 \tau ^2-1\right) n^2
\right. \right.\nn \\ && \left.\left.
+48 n' \left(3 \tau ^2-1\right) n+48 t \tau ^2+41 \tau ^2+24 n'^2 \left(3 \tau ^2-1\right)-15\right)\right)\Big) \ep+O\left(\ep^{3/2}\right),
\nn \\
\label{phi2}
\phi^{\delta_{\rm TM}}&=&
-\frac{\Om_L}{t y^2+\Om_L}-\frac{(n+n') \Om_L \sqrt{t} \tau  y^2 }{\left(t y^2+\Om_L\right)^2}\sqrt{\ep}
 \\ &&
+\frac{1}{48 t \left(t y^2+\Om_L\right)^3}\Big(\Om_L \left(t^2 \left(24 \left(\tau ^2-1\right) n^2+48 n' \left(\tau ^2-1\right) n+48 t \tau ^2+7 \tau ^2
\right.\right. \nn \\ && \left.\left.
+24 n'^2 \left(\tau ^2-1\right)-9\right) y^4+2 \Om_L t \left(12 \left(3 \tau ^2-1\right) n^2
\right.\right. \nn \\ && \left.\left.
+24 n' \left(3 \tau ^2-1\right) n+24 t \tau ^2+13 \tau ^2+12 n'^2 \left(3 \tau ^2-1\right)-3\right) y^2
\right. \nn \\ && \left.
+\Om_L^2 \left(3-5 \tau ^2\right)\right) \Big)\ep
+O\left(\ep^{3/2}\right),
\nn \\
\label{phi3}
\phi^{\ep_{\rm TE}}&=&
\frac{\sqrt{\om_L^2+t^2}-t}{t+\sqrt{\om_L^2+t^2}}-\frac{2 \left((n+n') \om_L^2 \sqrt{t} \tau \right) }{\sqrt{\om_L^2+t^2} \left(t+\sqrt{\om_L^2+t^2}\right)^2}\sqrt{\ep}
 \\ &&
+\left(-\frac{\left(\om_L^2+t \left(-t \tau ^2-2 \sqrt{\om_L^2+t^2} \tau ^2+t\right)\right) \om_L^2}{2 \left(\om_L^2+t^2\right)^{3/2} \left(t+\sqrt{\om_L^2+t^2}\right)^2}
\right. \nn \\ && \left.
+\frac{1}{\left(\om_L^2+t^2\right)^2 \left(t+\sqrt{\om_L^2+t^2}\right)^3}   
\left(
\left(\left(\left(\tau ^2-1\right) \om_L^4
\right.\right.\right. \right. \nn \\ && \left.\left.\left.\left.
+t \left(\sqrt{\om_L^2+t^2} \left(3 \tau ^2-1\right)+t \left(5 \tau ^2-2\right)\right) \om_L^2+t^3 \left(t+\sqrt{\om_L^2+t^2}\right) \left(4 \tau ^2-1\right)\right) n^2
\right.\right. \right. \nn \\ && \left.\left.\left.
+2 n' \left(\left(\tau ^2-1\right) \om_L^4+t \left(\sqrt{\om_L^2+t^2} \left(3 \tau ^2-1\right)+t \left(5 \tau ^2-2\right)\right) \om_L^2
\right.\right.\right. \right. \nn \\ && \left.\left.\left.\left.
+t^3 \left(t+\sqrt{\om_L^2+t^2}\right) \left(4 \tau ^2-1\right)\right) n+2 t \left(\om_L^2+t^2\right) \left(\om_L^2+t \left(t+\sqrt{\om_L^2+t^2}\right)\right) \tau ^2
\right.\right. \right. \nn \\ && \left.\left.\left.
+n'^2 \left(\left(\tau ^2-1\right) \om_L^4+t \left(\sqrt{\om_L^2+t^2} \left(3 \tau ^2-1\right)+t \left(5 \tau ^2-2\right)\right) \om_L^2
\right.\right.\right. \right. \nn \\ && \left.\left.\left.\left.
+t^3 \left(t+\sqrt{\om_L^2+t^2}\right) \left(4 \tau ^2-1\right)\right)\right) \om_L^2
\right)
\right.\nn \\ &&\left.
+\frac{\left(\sqrt{\om_L^2+t^2}-t\right) \tau ^2}{2 \left(\om_L^2+t \left(t+\sqrt{\om_L^2+t^2}\right)\right)}+\frac{\left(\sqrt{\om_L^2+t^2}-t\right) \left(5 \tau ^2-3\right)}{48 t \left(t+\sqrt{\om_L^2+t^2}\right)}\right) \ep+O\left(\ep^{3/2}\right),
\nn \\
\label{phi4}
\phi^{\ep_{\rm TM}}&=&
\frac{t \left(\sqrt{\om_L^2+t^2}-t\right) y^2-\om_L^2}{\om_L^2+t \left(t+\sqrt{\om_L^2+t^2}\right) y^2}
-\frac{2 \left((n+n') \om_L^2 \sqrt{t} \tau  y^2 \left(\om_L^2+t^2 y^2\right)\right) }{\sqrt{\om_L^2+t^2} \left(\om_L^2+t \left(t+\sqrt{\om_L^2+t^2}\right) y^2\right)^2}\sqrt{\ep}
\\ &&
+\frac{1}{48 \left(\om_L^2+t \left(t+\sqrt{\om_L^2+t^2}\right) y^2\right)^3}
\nn \\ && \times
\left(\left(\frac{24 \om_L^2 \tau ^2 \left(\om_L^2+t^2 y^2\right) \left(\om_L^2+t \left(t+\sqrt{\om_L^2+t^2}\right) y^2\right) y^2}{\sqrt{\om_L^2+t^2}}
\right. \right. \nn \\ && \left.\left.
-\frac{1}{\left(\om_L^2+t^2\right)^{3/2}}\left(24 \om_L^2 \left(\om_L^2+t^2 y^2\right) \left(\left(1-2 \tau ^2\right) \om_L^4+t \left(\sqrt{\om_L^2+t^2} y^2
\right.\right.\right. \right.\right. \nn \\ && \left.\left.\left.\left. \left.
+t \left(-\left(2 y^2+3\right) \tau ^2+y^2+1\right)\right) \om_L^2-t^3 \left(t+\sqrt{\om_L^2+t^2}\right) \left(3 \tau ^2-1\right) y^2\right) y^2    \right)
\right. \right.  \nn \\ && \left. \left. 
+\frac{1}{\left(\om_L^2+t^2\right)^2}\left(48 \om_L^2 \left(\om_L^2+t^2 y^2\right) \left(\left(\left(\left(t y^2+3 \sqrt{\om_L^2+t^2}\right) \tau ^2-t y^2
\right.\right.\right. \right.\right.\right. \nn \\ && \left.\left.\left.\left. \left.\left.
-\sqrt{\om_L^2+t^2}\right) \om_L^4+t^2 \left(\left(5 t y^2+\sqrt{\om_L^2+t^2} \left(3 y^2+4\right)\right) \tau ^2-2 t y^2
\right.\right.\right. \right.\right.\right. \nn \\ && \left.\left.\left.\left. \left.\left.
-\sqrt{\om_L^2+t^2} \left(y^2+1\right)\right) \om_L^2+t^4 \left(t+\sqrt{\om_L^2+t^2}\right) \left(4 \tau ^2-1\right) y^2\right) n^2
\right.\right. \right.\right.\nn \\ && \left.\left.\left.\left. 
+2 n' \left(\left(\left(t y^2+3 \sqrt{\om_L^2+t^2}\right) \tau ^2-t y^2-\sqrt{\om_L^2+t^2}\right) \om_L^4
\right.\right. \right.\right.\right. \nn \\ && \left.\left.\left.\left. \left.
+t^2 \left(\left(5 t y^2+\sqrt{\om_L^2+t^2} \left(3 y^2+4\right)\right) \tau ^2-2 t y^2-\sqrt{\om_L^2+t^2} \left(y^2+1\right)\right) \om_L^2
\right.\right. \right.\right.\right. \nn \\ && \left.\left.\left.\left. \left.
+t^4 \left(t+\sqrt{\om_L^2+t^2}\right) \left(4 \tau ^2-1\right) y^2\right) n+2 t \left(\om_L^2+t^2\right) \tau ^2 \left(\left(t y^2+\sqrt{\om_L^2+t^2}\right) \om_L^2
\right.\right. \right.\right.\right. \nn \\ && \left.\left.\left.\left. \left.
+t^2 \left(t+\sqrt{\om_L^2+t^2}\right) y^2\right)+n'^2 \left(\left(\left(t y^2+3 \sqrt{\om_L^2+t^2}\right) \tau ^2-t y^2
\right.\right.\right. \right.\right.\right. \nn \\ && \left.\left.\left.\left. \left.\left.
-\sqrt{\om_L^2+t^2}\right) \om_L^4+t^2 \left(\left(5 t y^2+\sqrt{\om_L^2+t^2} \left(3 y^2+4\right)\right) \tau ^2-2 t y^2
\right.\right.\right. \right.\right.\right. \nn \\ && \left.\left.\left.\left. \left.\left.
-\sqrt{\om_L^2+t^2} \left(y^2+1\right)\right) \om_L^2+t^4 \left(t+\sqrt{\om_L^2+t^2}\right) \left(4 \tau ^2-1\right) y^2\right)\right) y^2\right)
\right.\right.\nn \\ && \left.\left.
-\frac{\left(5 \tau ^2-3\right)}{t} \left(\om_L^2+t \left(t-\sqrt{\om_L^2+t^2}\right) y^2\right) \left(\om_L^2+t \left(t+\sqrt{\om_L^2+t^2}\right) y^2\right)^2\right)\right) \ep
\nn \\ && \nn
+O\left(\ep^{3/2}\right).
\eea
These expansions were done machined, of course. 

Finally we collect the contributions resulting from the prefactor in \Ref{sp1} together with the expansion of the exponentials in $\A_{m+m'}$ and in $R^{\delta_{\rm TE}}$ and $R^{\delta_{\rm TM}}$ (which was denoted by $\eta_0$ in \Ref{67} and \Ref{68}). Up to a common factor these collect into
\bea \label{Apsi}
\psi&\equiv&
1+\frac{(n+n') \left(2 n^2-4 n' n+2 n'^2-4 t-1\right) \tau 
   }{2 \sqrt{t}}\sqrt{\ep}
   \nn \\ &&
   +\frac{1}{24 t}\left(-12 (n+n')^2 \left(n^2-2 n'
   n+n'^2-2 t\right) \tau ^2+3 \left(\left(5 \tau ^2-2\right) n^2
\right. \right.\nn \\ && \left. \left.   
+2 n' \left(5 \tau
   ^2-2\right) n+4 t \left(\tau ^2-1\right)+n'^2 \left(5 \tau ^2-2\right)\right)+2 \left(-7 \left(3
   \tau ^2-1\right) n^4
\right. \right.\nn \\ && \left. \left.   
+4 n' \left(3 \tau ^2-1\right) n^3+6 \left(\left(3 \tau ^2-1\right)
   n'^2+2 t \left(\tau ^2-1\right)\right) n^2
\right. \right.\nn \\ && \left. \left.   
   +4 n' \left(\left(3 \tau ^2-1\right)
   n'^2+6 t \left(\tau ^2-1\right)\right) n
\right. \right.\nn \\ && \left. \left.   
   +6 (n+n')^2 \left(n^2-2 n'
   n+n'^2-2 t\right)^2 \tau ^2+12 t^2 \tau ^2+12 n'^2 t \left(\tau ^2-1\right)
\right. \right.\nn \\ && \left. \left.   
   -7 n'^4
   \left(3 \tau ^2-1\right)\right)\right) \ep
   +O\left(\ep^{3/2}\right)
\eea
and are used in formula \Ref{A3}.
\setcounter{equation}{0}\renewcommand{\theequation}{C.\arabic{equation}}
\section*{Appendix C}
Here we collect the expressions for $a^{(1/2)}$ and $a^{(1)}$ in Eq.\Ref{A3}. These are the generalizations of Eqs. (B13) in \meins ~and for $\Om_L\to\infty$ and $\om_L\to\infty$ the old expressions reappear. The explicit formulas are quite lengthy,
\bea                                            \label{a12ddTE}
a^{(1/2)}_{(\delta\delta)_{\rm TE}}&=&
\frac{\tau}{2\sqrt{t}(\Om_L+t)}\Big(-\left(-2 (\Om_L+t) n^3+2 n' (\Om_L+t) n^2+\left(\Om_L \left(2
   n'^2+4 t-1\right)
\right.\right. \nn \\ && \left.\left. 
   +t \left(2 n'^2+4 t
   +5\right)\right) n+n' \left(\Om_L
   \left(-2 n'^2+4 t+3\right)
   +t \left(-2 n'^2+4 t+5\right)\right)\right) \Big)  ,  \\
a^{(1)}_{(\delta\delta)_{\rm TE}}&=&                       \label{a1ddTE}
\frac{1}{48t(\Om_L+t)^2}   
\Big(48 \tau ^2 t^3-72 n^2 t^2-72 n'^2 t^2+192 n^2 \tau ^2 t^2+192 n'^2 \tau ^2 t^2
\nn \\ &&
+192
   n n' \tau ^2 t^2+48 \Om_L \tau ^2 t^2+21 \tau ^2 t^2-48 n n' t^2-3 t^2+24
   n^2 \Om_L \tau ^2 t
\nn \\ &&
+264 n'^2 \Om_L \tau ^2 t-8 \left(18 n^2+5\right)
   \Om_L \tau ^2 t+48 n n' \Om_L \tau ^2 t+22 \Om_L \tau ^2 t
\nn \\ &&
-24 n^2
   \Om_L t-120 n'^2 \Om_L t-48 n n' \Om_L t+6 \Om_L t+48 n^2
   \Om_L^2-48 n'^2 \Om_L^2+9 \Om_L^2
\nn \\ &&
-72 n^2 \Om_L^2 \tau ^2+120 n'^2
   \Om_L^2 \tau ^2-48 n n' \Om_L^2 \tau ^2-15 \Om_L^2 \tau ^2
\nn \\ &&
+(\Om_L+t)^2
   \left(-84 \tau ^2 n^4+28 n^4+48 n' \tau ^2 n^3-16 n' n^3-24
   n'^2 n^2+72 n'^2 \tau ^2 n^2
\right.\nn \\ &&\left.
+48 t \tau ^2 n^2+30 \tau ^2 n^2-48
   t n^2-12 n^2-16 n'^3 n+48 n'^3 \tau ^2 n+60 n' \tau ^2
   n
\right.\nn \\ &&\left.
+96 n' t \tau ^2 n-24 n' n-96 n' t n+28 n'^4-12
   n'^2-84 n'^4 \tau ^2+30 n'^2 \tau ^2
\right.\nn \\ &&\left.
+48 t^2 \tau ^2+48 n'^2 t \tau ^2+24 t \tau
   ^2-48 n'^2 t-24 t\right)
\nn \\ &&
+(n+n') \left(-24 n (t-\Om_L) \left(\Om_L
   \left(2 n^2-4 n' n+2 n'^2-4 t-1\right)
\right.\right.\nn \\ &&\left.\left.
+\left(2 n^2-4 n'
   n+2 n'^2-4 t-3\right) t\right) \tau ^2-24 n' (\Om_L+t) \left(\Om_L \left(2
   n^2-4 n' n
\right.\right.\right.\nn \\ &&\left.\left.\left.
+2 n'^2-4 t-1\right)+\left(2 n^2-4 n' n+2
   n'^2-4 t-3\right) t\right) \tau ^2\right)
 \\ &&
+(n+n')^2 \left((\Om_L+t)^2 \left(24
   \left(n^2-2 n' n+n'^2-2 t\right)^2 \tau ^2
\right.\right.\nn \\ &&\left.\left.
   -24 \left(n^2-2 n'
   n+n'^2-2 t\right) \tau ^2\right)-24 \left(2 n^2-4 n' n+2 n'^2-4
   t-1\right) t (\Om_L+t) \tau ^2\right)\Big) \nn ,
\eea
\bea                                        \label{a12ddTM}
a^{(1/2)}_{(\delta\delta)_{\rm TM}}&=&
\frac{\tau}{2\sqrt{t}(\Om_L+t y^2)}
\Big(\Om_L \left(2 n^3-2 n' n^2-2 n'^2 n-4 t n-3 n+2
   n'^3+n'-4 n' t\right) \tau 
\nn \\ &&
-(n+n') t \left(-2 n^2+4 n'
   n-2 n'^2+4 t-3\right) \tau  y^2 \Big) ,
   \\
a^{(1)}_{(\delta\delta)_{\rm TM}}&=&                              \label{a1ddTM}
\frac{1}{48} \left(\frac{1}{t \left(t y^2+\Om_L\right)^2}\Big(24 (n+n') \left(n t y^2+n' t y^2-n
   \Om_L+n' \Om_L\right) 
\right.\nn \\ && \left.   
   \left(\left(2 n^2-4 n' n+2 n'^2-4
   t+1\right) t y^2
+\Om_L \left(2 n^2-4 n' n+2 n'^2-4
   t-1\right)\right) \tau ^2    \Big)
\right.\nn \\ && \left.   
   +\frac{1}{t
   y^2+\Om_L}\Big(  24 (n+n')^2 \left(2
   n^2-4 n' n+2 n'^2-4 t-1\right) y^2 \tau ^2   \Big)
\right.\nn \\ && \left.   
   +\frac{1}{t}\Big(
   2 \left(-12 (n+n')^2 \left(n^2-2 n'
   n+n'^2-2 t\right) \tau ^2+3 \left(\left(5 \tau ^2-2\right) n^2
\right.\right.\right.\nn \\ && \left.   \left.   \left.   
   +2 n' \left(5 \tau
   ^2-2\right) n+4 t \left(\tau ^2-1\right)+n'^2 \left(5 \tau ^2-2\right)\right)
\right.\right.\nn \\ && \left.   \left.   
   +2 \left(-7
   \left(3 \tau ^2-1\right) n^4+4 n' \left(3 \tau ^2-1\right) n^3+6 \left(\left(3 \tau
   ^2-1\right) n'^2
\right.\right.\right.\right.\nn \\ && \left.   \left.   \left.    \left.   
   +2 t \left(\tau ^2-1\right)\right) n^2+4 n' \left(\left(3 \tau
   ^2-1\right) n'^2+6 t \left(\tau ^2-1\right)\right) n
\right.\right.\right.\nn \\ && \left.   \left.   \left.    
   +6 (n+n')^2
   \left(n^2-2 n' n+n'^2-2 t\right)^2 \tau ^2+12 t^2 \tau ^2
\right.\right.\right.\nn \\ && \left.   \left.   \left.    
   +12 n'^2 t
   \left(\tau ^2-1\right)-7 n'^4 \left(3 \tau ^2-1\right)\right)\right)   \Big)
\right.\nn \\ && \left.
   -\frac{1}{t \left(t   y^2+\Om_L\right)^2}   \Big(t^2 \left(24 \left(\tau
   ^2-1\right) n^2+48 n' \left(\tau ^2-1\right) n+48 t \tau ^2
\right.\right.\nn \\ && \left.   \left.   
   +7 \tau ^2+24 n'^2
   \left(\tau ^2-1\right)-9\right) y^4+2 \Om_L t \left(12 \left(3 \tau ^2-1\right) n^2
\right.\right.\nn \\ && \left.   \left.
+24
   n' \left(3 \tau ^2-1\right) n+24 t \tau ^2+13 \tau ^2+12 n'^2 \left(3 \tau
   ^2-1\right)-3\right) y^2+\Om_L^2 \left(3-5 \tau ^2\right)   \Big)
\right.\nn \\ && \left.  
   +\frac{1}{t \left(t y^2+\Om_L\right)^2} \Big(
   4 \left(t^2 \left(-6 \left(3 \tau ^2-2\right) n^2+12 n'
   \tau ^2 n+7 \tau ^2
\right.\right.\right.\nn \\ && \left.   \left. \left.   
   -6 n'^2 \left(3 \tau ^2-2\right)-9\right) y^4-2 \Om_L t \left(6
   \left(3 \tau ^2-2\right) n'^2+\left(18 n^2-7\right) \tau ^2+9\right) y^2
\right.\right.\nn \\ && \left.   \left.
   +\Om_L^2
   \left(6 \left(5 \tau ^2-2\right) n^2-12 n' \tau ^2 n+7 \tau ^2-6 n'^2 \left(3
   \tau ^2-2\right)-9\right)\right)\Big)\right) ,
\eea
\bea                                       \label{a12epdTE}
a^{(1/2)}_{(\ep\delta)_{\rm TE}}&=&-\frac{(n (t-\Om_L)+n' (\Om_L+t)) \tau }{\sqrt{t}
   (\Om_L+t)}
\nn \\ &&
+\frac{(n+n') \left(2 n^2-4 n' n+2 n'^2-4
   t-1\right) \tau }{2 \sqrt{t}}
   -\frac{2 (n+n') \sqrt{t} \tau }{\sqrt{\om_L^2+t^2}}   , \\
a^{(1)}_{(\ep\delta)_{\rm TE}}&=&                              \label{a1epdTE}
-\frac{(n+n') (n (t-\Om_L)+n' (\Om_L+t)) \left(2 n^2-4 n'
   n+2 n'^2-4 t-\frac{4 t}{\sqrt{\om_L^2+t^2}}-1\right) \tau ^2}{2 t
   (\Om_L+t)}
\nn \\ &&
   +\frac{(n+n')^2 \left(-2 n^2+4 n' n-2 n'^2+4
   t+1\right) \tau ^2}{\sqrt{\om_L^2+t^2}}
\nn \\ &&
+\frac{1}{12 t (\Om_L+t)^2}\Big(\left(-6 \left(3 \tau ^2-2\right) n^2-12 n'
   \tau ^2 n-5 \tau ^2
\right.\nn \\ && \left.    
   +6 n'^2 \left(5 \tau ^2-2\right)+3\right) \Om_L^2+2 t \left(6 \left(5
   \tau ^2-2\right) n'^2-\left(18 n^2+5\right) \tau ^2+3\right) \Om_L
\nn \\ &&   
   +t^2 \left(6 \left(5
   \tau ^2-2\right) n^2+12 n' \tau ^2 n-5 \tau ^2+6 n'^2 \left(5 \tau
   ^2-2\right)+3\right)\Big)
\nn \\ &&
+\frac{1}{24 t} \Big(-12 (n+n')^2 \left(n^2-2 n'
   n+n'^2-2 t\right) \tau ^2
\nn \\ &&
   +3 \left(\left(5 \tau ^2-2\right) n^2+2 n' \left(5 \tau
   ^2-2\right) n+4 t \left(\tau ^2-1\right)+n'^2 \left(5 \tau ^2-2\right)\right)
\nn \\ &&
+2 \left(-7
   \left(3 \tau ^2-1\right) n^4+4 n' \left(3 \tau ^2-1\right) n^3+6 \left(\left(3 \tau
   ^2-1\right) n'^2
\right.\right.\nn \\ && \left.\left.     
   +2 t \left(\tau ^2-1\right)\right) n^2+4 n' \left(\left(3 \tau
   ^2-1\right) n'^2+6 t \left(\tau ^2-1\right)\right) n
\right.\nn \\ && \left.        
   +6 (n+n')^2
   \left(n^2-2 n' n+n'^2-2 t\right)^2 \tau ^2+12 t^2 \tau ^2+12 n'^2 t
   \left(\tau ^2-1\right)
\right.\nn \\ && \left.   
   -7 n'^4 \left(3 \tau ^2-1\right)\right)   \Big)
\nn \\ &&
+\frac{1}{48 t \left(\om_L^2+t^2\right)^{3/2}}\Big(\left(96 t^2 \tau ^2+24 \left(2 n^2+4 n' n+2 n'^2+1\right) t \left(\tau
   ^2-1\right)
\right.\nn \\ && \left.     
   +\sqrt{\om_L^2+t^2} \left(5 \tau ^2-3\right)\right) \om_L^2
\nn \\ &&   
   +t^2 \left(96 t^2 \tau ^2+24
   \left(2 n^2+4 n' n+2 n'^2+1\right) t \left(2 \tau
   ^2-1\right)
\right.\nn \\ && \left.   
   +\sqrt{\om_L^2+t^2} \left(\left(96 n^2+192 n' n+96
   n'^2+53\right) \tau ^2-3\right)\right)\Bigg) ,
\eea

\bea                                       \label{a12epdTM}
a^{(1/2)}_{(\ep\delta)_{\rm TM}}&=&
\frac{2 (n+n') \sqrt{t} \tau  \left(\om_L^2+t^2 y^2\right)
   y^2}{\sqrt{\om_L^2+t^2} \left(\om_L^2-t^2 y^2
   \left(y^2-2\right)\right)}
+\frac{(n+n') \left(2 n^2-4 n' n+2
   n'^2-4 t-1\right) \tau }{2 \sqrt{t}}
\nn \\ &&
+\frac{\tau  \left(n t y^2+n' t
   y^2-n \Om_L+n' \Om_L\right)}{\sqrt{t} \left(t y^2+\Om_L\right)}    , \\
a^{(1)}_{(\ep\delta)_{\rm TM}}&=&                              \label{a1epdTM}
-\frac{ (n+n')^2 \left(-2 n^2+4 n' n-2 n'^2+4
   t+1\right) \tau ^2 \left(\om_L^2+t^2 y^2\right) y^2}{\sqrt{\om_L^2+t^2}
   \left(\om_L^2-t^2 y^2 \left(y^2-2\right)\right)}
\nn \\ &&
+\frac{1}{24t}   \Big(  -12 (n+n')^2
   \left(n^2-2 n' n+n'^2-2 t\right) \tau ^2+3 \left(\left(5 \tau ^2-2\right)
   n^2
\right. \nn \\ && \left.   
   +2 n' \left(5 \tau ^2-2\right) n+4 t \left(\tau ^2-1\right)+n'^2 \left(5 \tau
   ^2-2\right)\right)
\nn \\ &&  
   +2 \left(-7 \left(3 \tau ^2-1\right) n^4+4 n' \left(3 \tau ^2-1\right)
   n^3+6 \left(\left(3 \tau ^2-1\right) n'^2
\right.\right. \nn \\ && \left. \left.
   +2 t \left(\tau ^2-1\right)\right) n^2+4
   n' \left(\left(3 \tau ^2-1\right) n'^2+6 t \left(\tau ^2-1\right)\right) n
\right. \nn \\ && \left.    
   +6
   (n+n')^2 \left(n^2-2 n' n+n'^2-2 t\right)^2 \tau ^2+12 t^2 \tau
   ^2
\right. \nn \\ && \left.
   +12 n'^2 t \left(\tau ^2-1\right)-7 n'^4 \left(3 \tau ^2-1\right)\right)  \Big)
\nn \\ &&
+\frac{1}{24t \left(t
   y^2+\Om_L\right)^2}      \Big(2 \left(t^2
   \left(-6 \left(3 \tau ^2-2\right) n^2+12 n' \tau ^2 n+7 \tau ^2
\right.\right. \nn \\ && \left. \left.
   -6 n'^2 \left(3
   \tau ^2-2\right)-9\right) y^4-2 \Om_L t \left(6 \left(3 \tau ^2-2\right) n'^2
\right.\right. \nn \\ && \left. \left.
   +\left(18
   n^2-7\right) \tau ^2+9\right) y^2+\Om_L^2 \left(6 \left(5 \tau ^2-2\right) n^2-12
   n' \tau ^2 n+7 \tau ^2
\right.\right. \nn \\ && \left. \left.
   -6 n'^2 \left(3 \tau ^2-2\right)-9\right)\right)   \Big)
%
\nn \\ &&
+\frac{1}{2 t \left(t y^2+\Om_L\right)}(n+n') \tau ^2 \left(n t y^2+n' t y^2-n
   \Om_L+n' \Om_L\right) 
\nn \\ &&
   \left(2 n^2-4 n' n+2 n'^2-4 t+\frac{4 t
   y^2 \left(\om_L^2+t^2 y^2\right)}{\sqrt{\om_L^2+t^2} \left(\om_L^2-t^2
   y^2 \left(y^2-2\right)\right)}-1\right)
\nn \\ &&
+\frac{1}{48 t
   \left(\om_L^2+t^2\right)^{3/2} \left(\om_L^2-t^2 y^2 \left(y^2-2\right)\right)^2}
\Big( 
\left(\left(-96 t^2 y^2
\right.\right. \nn \\ && \left. \left.
-72 \left(2 n^2+4 n' n+2 n'^2+1\right) t
   y^2+5 \sqrt{\om_L^2+t^2}\right) \tau ^2
\right. \nn \\ && \left.
   +24 \left(2 n^2+4 n' n+2
   n'^2+1\right) t y^2-3 \sqrt{\om_L^2+t^2}\right) \om_L^6
\nn \\ &&
   +t^2 \left(\left(24 \left(2
   n^2+4 n' n+2 n'^2+1\right) t \left(y^4-9 y^2-4\right)
   y^2
\right. \right.\nn \\ && \left.\left.
   +96 t^2 \left(y^4-3 y^2-1\right) y^2
\right. \right.\nn \\ && \left.\left.
   +\sqrt{\om_L^2+t^2} \left(2
   \left(48 n^2+96 n' n+48 n'^2+19\right) y^4+20
   y^2+5\right)\right) \tau ^2
\right.\nn \\ && \left.
   +3 \left(\sqrt{\om_L^2+t^2} \left(2 y^4-4
   y^2-1\right)
\right. \right.\nn \\ && \left.\left.   
   -8 \left(2 n^2+4 n' n+2 n'^2+1\right) t y^2
   \left(y^4-3 y^2-1\right)\right)\right) \om_L^4
\nn \\ &&
   +t^4 y^2 \left(\tau ^2 \left(24
   \left(2 n^2+4 n' n+2 n'^2+1\right) t \left(y^4-4 y^2-12\right)
   y^2
\right. \right.\nn \\ && \left.\left.
+96 t^2 \left(y^4-y^2-3\right) y^2+\sqrt{\om_L^2+t^2} \left(5
   y^6+4 \left(48 n^2+96 n' n
\right. \right.\right. \right.\nn \\ && \left.\left.\left.\left.
   +48 n'^2+19\right) y^4+10
   y^2+20\right)\right)
\right.\nn \\ &&\left.
   -3 \left(8 \left(2 n^2+4 n' n+2 n'^2+1\right) t
   \left(y^4-y^2-3\right) y^2
\right. \right.\nn \\ && \left.\left.   
   +\sqrt{\om_L^2+t^2} \left(y^6-4 y^4+2
   y^2+4\right)\right)\right) \om_L^2
\nn \\ && 
+t^6 y^4 \left(\tau ^2 \left(48 \left(2 n^2+4
   n' n+2 n'^2+1\right) t \left(y^2-4\right) y^2
\right. \right.\nn \\ && \left.\left.
   +96 t^2
   \left(y^2-2\right) y^2+\sqrt{\om_L^2+t^2} \left(\left(96 n^2+192 n'
   n+96 n'^2+53\right) y^4
\right.\right. \right.\nn \\ && \left.\left.\left.
   -20 y^2+20\right)\right)-3 \left(y^2-2\right)
   \left(8 \left(2 n^2+4 n' n+2 n'^2+1\right) t
   y^2
\right. \right.\nn \\ && \left.\left.
   +\sqrt{\om_L^2+t^2} \left(y^2-2\right)\right)\right) \Bigg) .
\eea
%
\setcounter{equation}{0}\renewcommand{\theequation}{D.\arabic{equation}}
\section*{Appendix D}
In this appendix we collect the formulas for the functions $f_1$, introduced in Eq.\Ref{Efg} describing the contributions beyond PFA. For the considered models these functions read
\bea\label{f0g0D}
f_1^{(\delta\delta)_{\rm TE}}(\Om L)&=&\frac{480\sqrt{2}}{\pi^{9/2}}
\ \frac{36}{7} \ 
\int_0^\infty dt \ t^{3/2} \ 
g^{(\delta\delta)_{\rm TE}}(\Om L,t) \ ,
\nn \\ 
f_1^{(\delta\delta)_{\rm TM}}(\Om L)&=&\frac{480\sqrt{2}}{\pi^{9/2}}
\ \frac{1}{\frac{7}{36}-\frac{40}{3\pi^2}} \ 
\int_0^\infty dt \ t^{3/2} \int_0^1 dy \ 
g^{(\delta\delta)_{\rm TM}}(\Om L,t,y) \ ,
\nn \\
f_1^{(\ep\delta)_{\rm TE}}(\Om L,\om_pL)&=&\frac{480\sqrt{2}}{\pi^{9/2}}
\ \frac{36}{7} \ 
\int_0^\infty dt \ t^{3/2}  \ 
g^{(\ep\delta)_{\rm TE}}(\Om L,t) \ ,
\nn \\
f_1^{(\ep\delta)_{\rm TM}}(\Om L,\om_pL)&=&\frac{480\sqrt{2}}{\pi^{9/2}}
\ \frac{1}{\frac{7}{36}-\frac{40}{3\pi^2}} \ 
\int_0^\infty dt \ t^{3/2} \int_0^1 dy \ 
g^{(\ep\delta)_{\rm TM}}(\Om L,t,y) \ ,
\nn
\eea
where the functions $g$ for the $(\delta\delta)$-model are given by
\bea\label{D1}
&&g^{(\delta\delta)_{\rm TE}}(\Om L,t)=
\nn \\ &&
\left(16 t^4+32 \Om_L t^3+32 t^3+16 \Om_L^2 t^2+32 \Om_L t^2+16 t^2\right) \text{Li}_{-\frac{3}{2}}\left(\frac{e^{-2 t}
   \Om_L^2}{(\Om_L+t)^2}\right)
\nn \\ &&
+\left(-40 t^3-80 \Om_L t^2-24 t^2-40 \Om_L^2 t-40 \Om_L t\right)
   \text{Li}_{-\frac{1}{2}}\left(\frac{e^{-2 t} \Om_L^2}{(\Om_L+t)^2}\right)
\nn \\ &&
+\left(32 t^4+64 \Om_L t^3+16 t^3+32 \Om_L^2 t^2+16
   \Om_L t^2+11 t^2+30 \Om_L t+15 \Om_L^2\right) \text{Li}_{\frac{1}{2}}\left(\frac{e^{-2 t}
   \Om_L^2}{(\Om_L+t)^2}\right)
\nn \\ &&
+\left(-8 t^3-16 \Om_L t^2+12 t^2-8 \Om_L^2 t+16 \Om_L t\right)
   \text{Li}_{\frac{3}{2}}\left(\frac{e^{-2 t} \Om_L^2}{(\Om_L+t)^2}\right)
\nn \\ &&
+\left(-3 \Om_L^2-6 t \Om_L-3 t^2\right)
   \text{Li}_{\frac{5}{2}}\left(\frac{e^{-2 t} \Om_L^2}{(\Om_L+t)^2}\right) \ ,
\\ 
&&g^{(\delta\delta)_{\rm TM}}(\Om L,t,y)=               \label{D2}
\nn \\ &&
-16 t^2 \left(y^2-1\right) \left((t-1)
   y^2+\Om_L\right)^2 \text{Li}_{-\frac{3}{2}}\left(\frac{e^{-2 t} \Om_L^2}{\left(t y^2+\Om_L\right)^2}\right)
\nn \\ &&
   -8 t \left(t \left(-3 y^2+t \left(y^2+3\right)-1\right) y^4
\right. \nn \\ && \left.
   +\Om_L \left(-5 y^2+2 t \left(y^2+3\right)+1\right)
   y^2+\Om_L^2 \left(y^2+3\right)\right) \text{Li}_{-\frac{1}{2}}\left(\frac{e^{-2 t} \Om_L^2}{\left(t
   y^2+\Om_L\right)^2}\right)
\nn \\ &&
   +\left(-32 t^4 y^6+16 t^3 y^6-23 t^2 y^6+32 t^4 y^4-64 \Om_L t^3 y^4-16 t^3 y^4+16 \Om_L t^2 y^4-49 t^2
   y^4
\right. \nn \\ && \left.
   -54 \Om_L t y^4+64 \Om_L t^3 y^2-27 \Om_L^2 y^2-32 \Om_L^2 t^2 y^2-16 \Om_L t^2 y^2-90 \Om_L t y^2
\right. \nn \\ && \left.
-45
   \Om_L^2+32 \Om_L^2 t^2\right) \text{Li}_{\frac{1}{2}}\left(\frac{e^{-2 t} \Om_L^2}{\left(t
   y^2+\Om_L\right)^2}\right)
\nn \\ &&
   +\left(-16 t^3 y^6-32 \Om_L t^2 y^4-8 t^2 y^4-4 \Om_L t y^4-16 \Om_L^2 t y^2-4 \Om_L t
   y^2\right) \text{Li}_{\frac{3}{2}}\left(\frac{e^{-2 t} \Om_L^2}{\left(t y^2+\Om_L\right)^2}\right)
\nn \\ &&
+\left(-6 t^2 y^6-12 \Om_L t
   y^4-6 \Om_L^2 y^2\right) \text{Li}_{\frac{5}{2}}\left(\frac{e^{-2 t} \Om_L^2}{\left(t y^2+\Om_L\right)^2}\right) \ .
\eea
The corresponding function for the $(\ep\delta)$-model have a similar structure but they are too lengthy as to be displayed here. 

\bibliographystyle{unsrt}\bibliography{../../../Literatur/Bordag,../../../Literatur/articoli}
\end{document}